\documentclass[fleqn,usenatbib]{mnras}

\usepackage{newtxtext,newtxmath}

\usepackage[T1]{fontenc}

\DeclareRobustCommand{\VAN}[3]{#2}
\let\VANthebibliography\thebibliography
\def\thebibliography{\DeclareRobustCommand{\VAN}[3]{##3}\VANthebibliography}

\usepackage{graphicx}	
\usepackage{amsmath}	
\usepackage{pdflscape}
\usepackage{graphicx, bm, xcolor, colortbl}
\definecolor{nocloud}{RGB}{0,33,80}
\definecolor{CObrightH2dark}{RGB}{142,81,84}
\definecolor{H2brightCOdark}{RGB}{242,214,41}
\definecolor{H2brightCObright}{RGB}{210,168,144}

\usepackage{bm}
\usepackage{array}

\newcommand{\be}{\begin{equation}}
\newcommand{\ee}{\end{equation}}
\newcommand{\bea}{\begin{eqnarray}}
\newcommand{\eea}{\end{eqnarray}}
\newcommand{\dd}{\mathrm{d}}

\usepackage{scalerel,tikz}
\usetikzlibrary{svg.path}
\definecolor{orcidlogocol}{HTML}{A6CE39}
\tikzset{orcidlogo/.pic={
 \fill[orcidlogocol] svg{M256,128c0,70.7-57.3,128-128,128C57.3,256,0,198.7,0,128C0,57.3,57.3,0,128,0C198.7,0,256,57.3,256,128z};
 \fill[white] svg{M86.3,186.2H70.9V79.1h15.4v48.4V186.2z}
 svg{M108.9,79.1h41.6c39.6,0,57,28.3,57,53.6c0,27.5-21.5,53.6-56.8,53.6h-41.8V79.1z M124.3,172.4h24.5c34.9,0,42.9-26.5,42.9-39.7c0-21.5-13.7-39.7-43.7-39.7h-23.7V172.4z}
 svg{M88.7,56.8c0,5.5-4.5,10.1-10.1,10.1c-5.6,0-10.1-4.6-10.1-10.1c0-5.6,4.5-10.1,10.1-10.1C84.2,46.7,88.7,51.3,88.7,56.8z};
}}
\newcommand\orcidicon[1]{\href{https://orcid.org/#1}{\mbox{\scalerel*{
\begin{tikzpicture}[yscale=-1,transform shape]
\pic{orcidlogo};
\end{tikzpicture}
}{|}}}}

\title[Long-lived clouds, short-lived ${\rm H_2}$ molecules]{
Clouds of Theseus:
long-lived molecular clouds are composed of short-lived ${\rm H_2}$ molecules}

\author[Jeffreson, Semenov \& Krumholz]{
	Sarah M.~R.~Jeffreson$^{1\orcidicon{0000-0002-4232-0200}}$, Vadim A. Semenov$^{1\orcidicon{0000-0002-6648-7136}}$ and Mark R.~Krumholz$^{2,3\orcidicon{0000-0003-3893-854X}}$
	\\
$^{1}$ Center for Astrophysics, Harvard \& Smithsonian, 60 Garden St, Cambridge, MA 02138, USA \\
$^{2}$ Research School of Astronomy and Astrophysics, Australian National University, Canberra, ACT 2611 Australia \\
$^{3}$ Australian Research Council Centre of Excellence for All Sky Astrophysics in 3 Dimensions (ASTRO 3D), Australia \\
}

\date{Accepted XXX. Received XXX; in original form XXX}

\pubyear{2021}

\begin{document}
\label{firstpage}
\pagerange{\pageref{firstpage}--\pageref{lastpage}}
\maketitle

\begin{abstract}
We use passive gas tracer particles in an {\sc Arepo} simulation of a dwarf spiral galaxy to relate the Lagrangian evolution of star-forming gas parcels and their ${\rm H_2}$ molecules to the evolution of their host giant molecular clouds. We find that the median chemical lifetime of ${\rm H_2}$ is just $4~{\rm Myr}$, independent of the lifetime of its host molecular cloud, which may vary from 1 to 90~Myr, with a substantial portion of all star formation in the galaxy occurring in relatively long-lived clouds. The rapid ejection of gas from around young massive stars by early stellar feedback is responsible for this short ${\rm H_2}$ survival time, driving down the density of the surrounding gas, so that its ${\rm H_2}$ molecules are dissociated by the interstellar radiation field. This ejection of gas from the ${\rm H_2}$-dominated state is balanced by the constant accretion of new gas from the galactic environment, constituting a `competition model' for molecular cloud evolution. Gas ejection occurs at a rate that is proportional to the molecular cloud mass, so that the cloud lifetime is determined by the accretion rate, which may be as high as $4 \times 10^4~{\rm M}_\odot {\rm Myr}^{-1}$ in the longest-lived clouds. Our findings therefore resolve the conflict between observations of rapid gas ejection around young massive stars and observations of long-lived molecular clouds in galaxies, that often survive up to several tens of Myr. We show that the fastest-accreting, longest-lived, highest-mass clouds drive supernova clustering on sub-cloud scales, which in turn is a key driver of galactic outflows.
\end{abstract}

\begin{keywords}
ISM:clouds -- ISM:evolution -- ISM: structure -- ISM: Galaxies -- Galaxies: star formation
\end{keywords}


\section{Introduction} \label{Sec::Introduction}
In recent years, a new, dynamical picture of star formation has emerged. Numerical simulations of disk galaxies have demonstrated that Lagrangian parcels of gas in the interstellar medium undergo constant and rapid cycles of compression into, and disruption out of, a high-density `star-forming state' \citep{Semenov17,Semenov18,Shin22}. The average time spent by gas in the star-forming state ($\sim 5$-$15$~Myr) is much shorter than the time spent outside the star-forming state ($\ga 100$~Myr). This picture can explain several of the key observed attributes of star formation in disc galaxies. The fact that galaxies convert only a small fraction of their gas to stars per galactic rotation~\citep[e.g.][]{Kennicutt98,Wyder09,Daddi10} is consistent with the small fraction of time spent by gas in the star-forming state. The observed spread of star-formation rates (SFRs) at a given molecular gas surface density found when galaxies are observed at high resolution~\citep{Onodera10a, Schruba11}, and the relatively poor correlation between molecular masses and star formation rate indicators for clouds in the Milky Way \citep{Mooney88a, Lee16}, reflect the fact that any particular parcel of H$_2$ rich gas may have just started forming stars (in which case its luminosity per unit gas mass in SFR tracers such as H$\alpha$ or infrared emission will be low) or may have just completed a star formation cycle and left the star-forming state (in which case the luminosity per unit mass will be high). The spatial decorrelation of molecular gas and young stars on small scales ($\lesssim100~{\rm pc}$) relative to galactic scales~\citep[$\sim 1$~kpc,][]{Schruba10,Kruijssen2019} and the rapidity with which star clusters become optically revealed \citep{Hollyhead15a, Sokal16a} are consistent with the disruption of dense, star-forming gas on short ($<5$~Myr) time-scales by the radiation and thermal pressure from young, massive stars (pre-supernova stellar feedback). Finally, the near-proportionality of the star formation rate surface density $\Sigma_{\rm SFR}$ and the molecular gas surface density $\Sigma_{\rm H_2}$ observed on kpc scales \citep[e.g.,][]{WongBlitz2002,Bigiel08,Leroy+13} can also be explained by a self-regulation process, in which the density distribution of molecular gas is shaped by the stars it forms~\citep{Semenov19}.

However, the connection between this Lagrangian picture of star formation and the properties of observable star-forming regions or `giant molecular clouds' remains unclear. Molecular clouds are observed to have a large range of masses, sizes and densities, spanning over two orders of magnitude in Milky Way-like disc galaxies~\citep[e.g.][]{Sun18,Colombo+19}. These star-forming regions are \textit{not} universally destroyed on time-scales of $5$-$15$~Myr. Both observations~\citep[e.g.][]{ScovilleHersh1979,Engargiola03,Blitz2007,Murray11,Corbelli17} and simulations~\citep[e.g.][]{DobbsPringle13,Grisdale19, 2021MNRAS.505.1678J} show a large range of survival times for giant molecular clouds, ranging over two orders of magnitude, from $1$ to $>100$~Myr. While short-lived ($\sim 10$~Myr), low-mass ($\sim 10^4~{\rm M}_\odot$) molecular clouds dominate by number, a substantial fraction of molecular mass (and thus galactic star formation) resides in the most massive ($\ga 10^6~{\rm M}_\odot$) molecular clouds~\citep[e.g.][]{Murray11,Miville-Deschenes17,Faesi18}, which tend to live for longer periods of time~\citep[e.g.][]{2021MNRAS.505.1678J}. As such, the mass-weighted molecular cloud lifetime is typically up to a factor of $10$ longer than the star-forming lifetime of a Lagrangian gas parcel. We may therefore ask: \textbf{how do we reconcile the short Lagrangian star formation time-scale with the long lifetimes of observable molecular clouds?}

This question is closely-related to another unanswered question in the field of star formation: \textbf{what is the chemical lifetime of molecular hydrogen and how does it relate to the molecular cloud lifetime?} Based on observations of massive molecular clouds in the low-density inter-arm regions of spiral galaxies, \cite{ScovilleHersh1979} and~\cite{Koda09,Koda16a} argue that the survival time of hydrogen molecules must be comparable to the inter-arm crossing time. Given that the long time-scale for the formation of new hydrogen molecules from the low-density, inter-arm gas prohibits the formation of new massive clouds in these regions, these authors argue that high-mass molecular agglomerations must form within the spiral arms, then later fragment into smaller (but still massive) molecular clouds as they transit into the inter-arm regions. However, this hypothesis is at odds with the short Lagrangian lifetimes of star-forming gas~\citep{Semenov17}: when cold, dense, star-forming gas parcels are ejected into a warm, diffuse non-star-forming state by stellar feedback on time-scales of $\sim 5$-$15$~Myr, their hydrogen molecules should also be dissociated by the interstellar radiation field.

The chemical lifetime of ${\rm H_2}$ has important consequences for interpreting the observed spatial distribution of young star clusters and dense molecular gas, and in particular their decorrelation on sub-kiloparsec scales~\citep[e.g.][]{Blitz2007,Kawamura09,2012ApJ...761...37M,Corbelli17,Schruba10,Kruijssen2019}. If the ${\rm H_2}$ survival time-scale is short, then the observed spatial decorrelation of young stars and ${\rm H_2}$ on sub-kiloparsec scales implies that stellar feedback rapidly dissociates molecular gas and destroys the star-forming gas in its vicinity, providing evidence for short lifetimes of star-forming regions. However, if the ${\rm H_2}$ survival time-scale is long, the same observations imply that the molecular gas is simply pushed away from young stars, where it can continue to form new stars.

In this work, we seek to reconcile the apparent contradiction between the Lagrangian and Eulerian views of molecular cloud lifetime, and to illuminate the relationship of cloud to chemical lifetimes. We use passive Lagrangian tracer particles in a chemodynamic simulation of an isolated disc galaxy in the moving-mesh code {\sc Arepo} to directly compare the star-forming time-scale of Lagrangian gas parcels, the chemical lifetime of molecular hydrogen, and the lifetimes of giant molecular clouds. We describe our numerical prescription in Section~\ref{Sec::sims}, and examine the distribution of molecular mass and star formation among short- and long-lived molecular clouds in Section~\ref{Sec::sim-props}. We compare our molecular cloud lifetimes to the Lagrangian star formation time-scale and the molecular hydrogen survival time in Section~\ref{Sec::results}, and use this result to develop a simple picture of molecular cloud evolution in Section~\ref{Sec::competition-model}. In Section~\ref{Sec::clustering}, we also show that the massive, long-lived molecular clouds formed in this picture account for the clustering of supernova feedback in our simulation, and are therefore the key drivers of galactic winds. We discuss our results in the context of the existing literature in Section~\ref{Sec::discussion}, and present our conclusions in Section~\ref{Sec::conclusions}.

\section{Simulation of a dwarf spiral galaxy} \label{Sec::sims}
Here we describe our simulation setup and methods.

\subsection{Initial conditions}
We simulate a dwarf flocculent spiral galaxy that is analagous in its gas and stellar mass distribution to the bulgeless nearby galaxy NGC300. The initial condition includes a dark matter halo at a mass resolution of $1.254\times 10^7~{\rm M}_\odot$, a stellar disc at a mass resolution of $3437~{\rm M}_\odot$, and a gas disc at a mass resolution of $859~{\rm M}_\odot$. The dark matter halo follows the profile of~\cite{Navarro97}, with a concentration parameter of $c=15.4$, a spin parameter of $\lambda=0.04$, a mass of $8.3\times 10^{10}~{\rm M}_\odot$ and a circular velocity of $V_{200} = 76~{\rm km~s}^{-1}$ at the virial radius. The stellar disc is of exponential form, with a mass of $1 \times 10^9~{\rm M}_\odot$, a scale-length of $1.39$~kpc, and a scale-height of $0.28$~kpc. The corresponding exponential gas disc extends beyond the stellar disc, with a mass of $2.2 \times 10^9~{\rm M_\odot}$ (giving a gas fraction of 68~per~cent) and a scale-length of $3.44~{\rm kpc}$.

\subsection{Hydrodynamics and chemistry}
The initial condition is evolved using the moving-mesh hydrodynamics code {\sc Arepo}~\citep{Springel10}. Within {\sc Arepo}, the gas dynamics is modelled on the unstructured moving mesh defined by the Voronoi tesselation about a discrete set of points, which move according to the local gas velocity. The gravitational acceleration vectors of the Voronoi gas cells, stellar particles and dark matter particles are computed using a hybrid TreePM gravity solver.

\begin{figure*}
	\includegraphics[width=\linewidth]{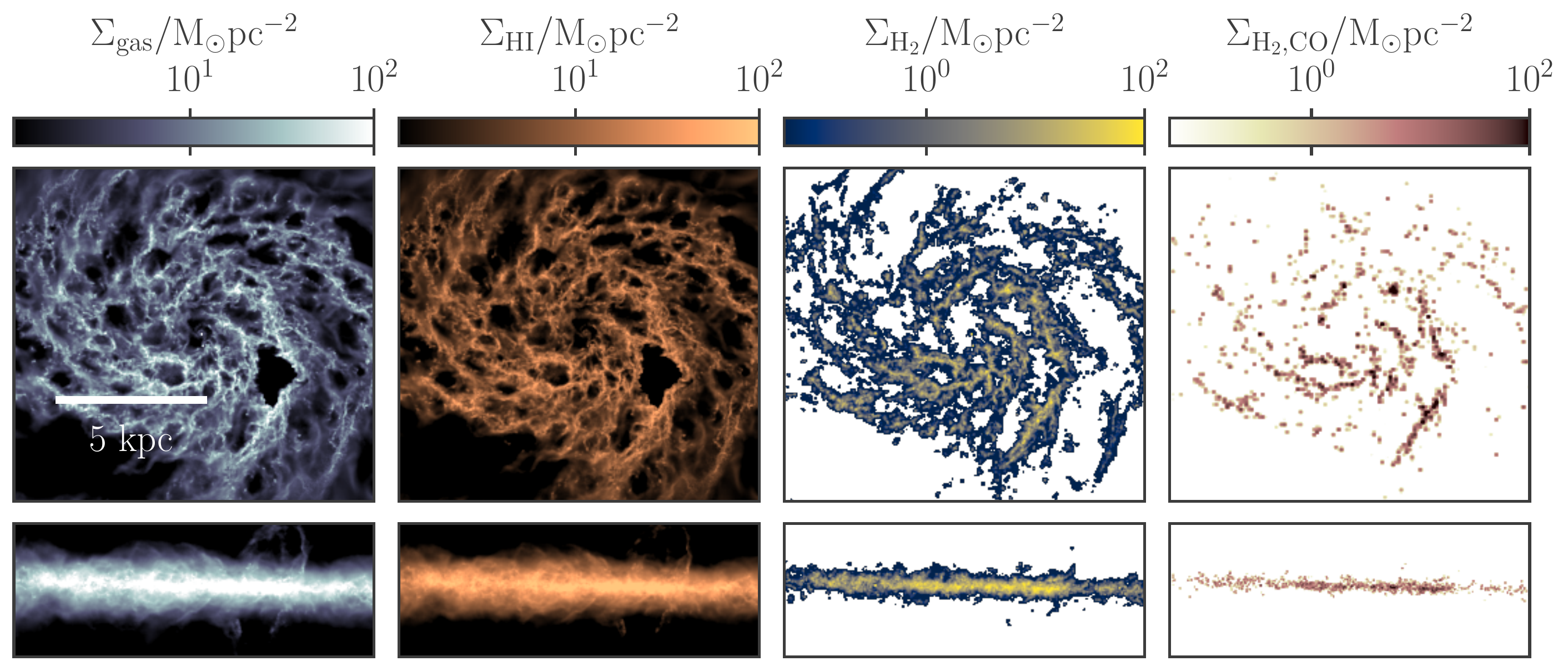}
	\caption{Column density maps of the total ($\Sigma_{\rm gas}$, left), atomic ($\Sigma_{\rm HI}$, centre-left), total molecular ($\Sigma_{\rm H_2}$, centre-right), and CO-luminous molecular ($\Sigma_{\rm H_2, CO}$ right) gas distribution for the simulated dwarf spiral galaxy, viewed perpendicular to (top panels) and across (lower panels) the galactic mid-plane, at a simulation time of $800$~Myr.}
	\label{Fig::morphology}
\end{figure*}

The temperature and chemical composition of the gas in our simulation is modelled using the simplified network of hydrogen, carbon and oxygen chemistry described in~\cite{NelsonLanger97} and in~\cite{GloverMacLow07a,GloverMacLow07b}. The fractional abundances of the chemical species ${\rm H}$, ${\rm H}_2$, ${\rm H}^+$, ${\rm He}$, ${\rm C}^+$, ${\rm CO}$, ${\rm O}$ and ${\rm e}^-$ are computed and tracked for each gas cell, and self-consistently coupled to the heating and cooling of the interstellar medium via the atomic and molecular cooling function of~\cite{Glover10}. The gas equilibrates to a state of thermal balance between line-emission cooling and heating due to the photo-electric emission from polycyclic aromatic hydrocarbons and dust grains, as they interact with the interstellar radiation field (ISRF) and with cosmic rays. We assign a value of $1.7$ Habing fields to the UV component of the ISRF~\citep{Mathis83}, a value of $3 \times 10^{-17}~{\rm s}^{-1}$ to the cosmic ray ionisation rate~\citep{2000A&A...358L..79V}, and we assume the solar value for the dust-to-gas ratio. We note that this results in a higher metallicity than the sub-solar value that is observed for NGC300~\citep{2009ApJ...700..309B}, but we demonstrate in Section~\ref{Sec::discussion::observations} that our simulation nevertheless produces a realistic interstellar medium structure, with a realistic distribution of molecular cloud properties.

We use the {\sc TreeCol} algorithm introduced by~\cite{Clark12} to model the dust- and self-shielding of molecular hydrogen from dissociation by the ISRF. This allows us to accurately model the non-equilibrium abundance of molecular hydrogen during the run-time of the simulation, and so to compute its value for each gas cell as a function of time.

\subsection{Tracer particles}
\label{Sec::sims::tracers}
We introduce passive tracer particles to the simulation following the Monte Carlo prescription of~\cite{2013MNRAS.435.1426G}, which allows us to track the Lagrangian mass flow and molecular fraction of gas as it moves through simulated GMCs, despite the fact that {\sc Arepo} is not a Lagrangian code. Via this prescription, tracer particles are moved along with the gas cells in the simulation, and are exchanged between gas cells according to a probability set by the mass flux between them. When a gas cell is converted to a star particle, the tracer particles associated with that gas cell are moved to the star particle with a probability set by the ratio of the stellar mass to the original gas cell mass. Similarly, tracer particles attached to star particles are transferred back to the gas reservoir according to the masses ejected in stellar winds and supernovae. As such, the mass of tracers in the gas and stellar reservoirs remains equal to the masses of these reservoirs throughout the simulation.

In this work, we assign one tracer particle to each gas cell in the initial condition, which sets an initial mass distribution for the tracer particles, equal to the initial mass distribution of gas cells. Since the cells are not of exactly equal mass initially, this means that the tracers are not uniform in mass either. However, we show in Appendix~\ref{App:tracer-error} that the distribution of effective tracer masses converges to a uniform value after after $<100$~Myr. We analyse the simulated disc only after it is in a state of dynamical equilibrium, between simulation times of $500$ and $800$~Myr. The effective tracer mass during this period is stable at a value of $\sim 450~{\rm M_\odot}$, corresponding to $1.9$ tracer particles per gas cell. In what follows, we also make sure to analyse only large populations of tracer particles to mitigate the effects of the Poisson noise: our results are derived for samples containing $10^2$ to $10^3$ distinct molecular clouds, the smallest of which sees the passage of $\sim 100$ tracer particles over its lifetime.

\subsection{Star formation and feedback}
The star formation efficiency $\epsilon_{\rm ff}$ per free-fall time of the gas in our simulations follows the parametrisation of~\cite{Padoan17}. These authors conduct high-resolution simulations of turbulent fragmentation and find that $\epsilon_{\rm ff}$ depends on the local virial parameter $\alpha_{\rm vir}$ of the gas, according to
\begin{equation} \label{Eqn::epsilon_ff}
\epsilon_{\rm ff} = 0.4 \exp{(-1.6 \alpha_{\rm vir}^{0.5})}.
\end{equation}
To use such a star formation recipe, simulations need a model for unresolved turbulent velocity dispersion, $\sigma$, to calculate $\alpha_{\rm vir} \propto \sigma^2/(\rho L^2)$. To this end, existing methods in galaxy formation simulations include explicit dynamic models for sub-grid turbulence \citep[e.g.,][]{BraunSchmidt15,Semenov16,KretschmerTeyssier20}, estimation of $\sigma$ from the gradients of resolved gas velocity \citep[e.g.,][]{Hopkins13sf}, or calculation of $\sigma$ over some resolved scale~\citep[e.g.,][]{Gensior20}.
As our simulations do not include a sub-grid model for turbulence, we follow the prescription of~\cite{Gensior20}. In brief, the scale over which $\alpha_{\rm vir}$ is calculated is computed using Sobolev approximation: $L=|\langle \rho_{\rm g} \rangle/\langle \nabla \rho_{\rm g} \rangle|$, where $\langle \nabla \rho_{\rm g} \rangle$ is the cubic spline kernel-weighted average of the gas volume density gradient, with respect to the radial distance from the central gas cell. The smoothing length of the cubic spline kernel is chosen to enclose the 32 nearest-neighbour cells. We refer the reader to~\cite{Gensior20} for a more detailed explanation. The star formation rate volume density of each gas cell in the simulation is therefore given by
\begin{equation}
\label{Eqn::starformation}
\frac{\dd \rho_{*,i}}{\dd t} = 
\begin{cases}
      \frac{\epsilon_{\rm ff} \rho_i}{t_{{\rm ff},i}}, & \rho_i \geq \rho_{\rm thresh}\mbox{ and } T_i \leq T_{\rm thresh} \\
      0, & \rho_i < \rho_{\rm thresh}\mbox{ or } T_i > T_{\rm thresh}\\
  \end{cases},
\end{equation}
where $t_{{\rm ff}, i} = \sqrt{3\pi/(32 G\rho_i)}$ is the local free-fall time-scale for the gas cell $i$ with a mass volume density of $\rho_i$, and $\epsilon_{\rm ff}$ is given by Equation~(\ref{Eqn::epsilon_ff}).

We set a lower limit of $\rho_{\rm thresh}/m_{\rm H}\mu = 100~{\rm cm}^{-3}$ on the volume density of hydrogen atoms above which star formation is allowed to occur, as well as an upper limit of $T_{\rm thresh} = 100$~K on the temperature; here $\mu\approx 1.4$ is the mean mass per H atom. The value of $\rho_{\rm thresh}$ is the density of Jeans-unstable (collapsing) gas at our mass resolution of $859~{\rm M}_\odot$ and at the temperature of $\sim 30$~K reached by the molecular gas in our simulation. For a spherical gas cell, the radius associated with this density threshold is $3$~pc, and so we employ the adaptive gravitational softening scheme in {\sc Arepo} with a minimum softening length of $6$~pc and a gradation of $1.5$ times the Voronoi gas cell diameter. We set the softening length of the stellar particles to the same value, and choose a softening length of $280$~pc for the dark matter particles, according to the convergence tests presented in~\cite{2003MNRAS.338...14P}. Because our simulations resolve the gas disc scale-height and the Toomre mass at all scales, the adaptive gravitational softening avoids the majority of artificial fragmentation at scales larger than the Jeans length~\citep{Nelson06}.

\begin{figure}
	\includegraphics[width=\linewidth]{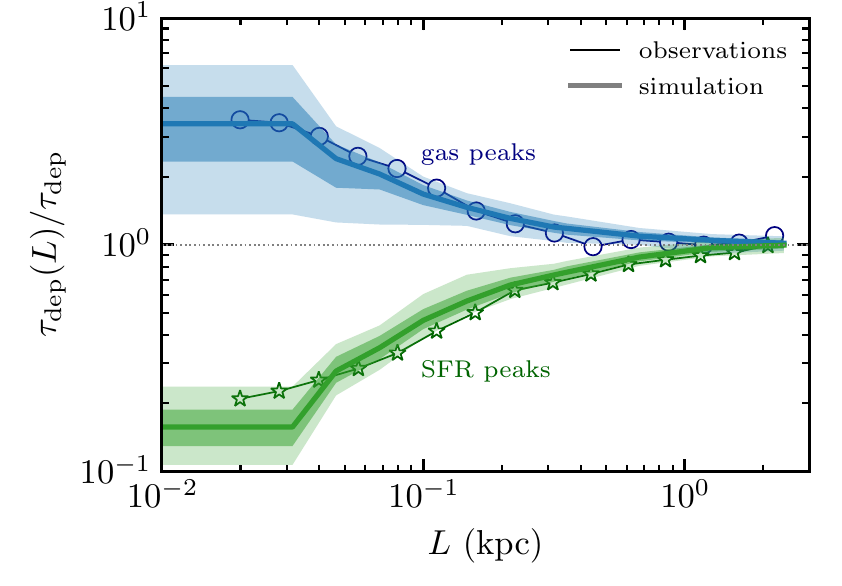}
	\caption{Our simulated dwarf spiral galaxy reproduces the spatial decorrelation observed for NGC300 between young stars and dense gas on the scales of molecular clouds—the so-called ``tuning fork diagram’’~\protect\citep[][thin lines with markers]{Kruijssen2019}. The branches of the tuning fork show the average bias of gas depletion times measured in the apertures of a given size ($x$-axis) placed on the peaks of dense gas (top branch) or young stars (lower branch). To quantify the snapshot-to-snapshot variation of the tuning fork in our simulation, the thick solid lines show medians and the shaded regions show 16--84 and 2.5–97.5 interpercentile ranges. The diagram is computed over the range of galactocentric radii of $R = 2\text{--}6$ kpc (see the text for details).}
	\label{Fig::tuning-fork}
\end{figure}

\begin{figure}
	\includegraphics[width=\linewidth]{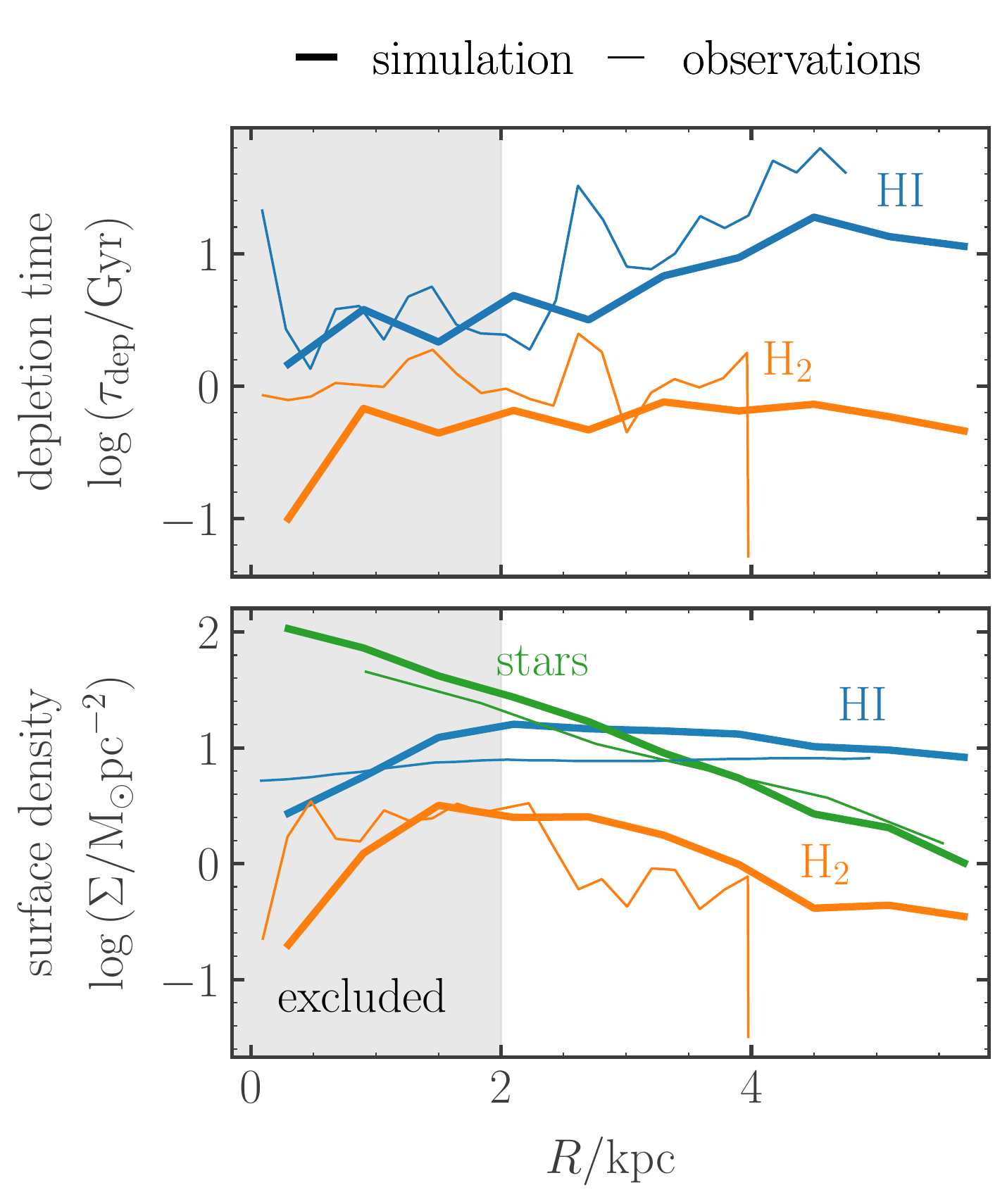}
	\caption{Our simulated dwarf spiral galaxy is similar to NGC300 in its gas and stellar surface densities and gas depletion times. \textit{Upper panel:} Molecular and atomic gas depletion times as a function of galactocentric radius (solid lines), measured within annuli of width $500$~pc. Thin lines show the values measured for NGC300 by~\protect\cite{Kruijssen2019}. \textit{Lower panel:} Stellar, molecular and atomic gas surface densities as a function of galactocentric radius, measured within the same annuli. Thin lines show the values measured for NGC300 by~\protect\cite{Kruijssen2019} and by~\protect\cite{2011MNRAS.410.2217W}.}
	\label{Fig::observables}
\end{figure}

The stellar feedback in our simulation is modelled via injection of energy and momentum by supernovae and pre-supernova HII regions.
To compute the number of supernovae generated by each star particle formed during the simulation, we assign a stellar population drawn stochastically from a~\cite{Chabrier03} initial stellar mass function (IMF), using the Stochastically Lighting Up Galaxies (SLUG) stellar population synthesis model~\citep{daSilva12,daSilva14,Krumholz15}. Within SLUG, the resulting stellar populations are evolved along Padova solar metallicity tracks~\citep{Fagotto94a,Fagotto94b,VazquezLeitherer05} during run-time, using {\sc Starburst99}-like spectral synthesis~\citep{Leitherer99}. This modelling provides the number of supernovae $N_{*, {\rm SN}}$ generated by each star particle during each simulation time-step, as well as the ionising luminosity of the cluster and the mass $\Delta m_*$ it has ejected.

We use the values of $N_{*, {\rm SN}}$ and $\Delta m_*$ for each star particle to compute the momentum and thermal energy injected by supernova explosions at each time-step. In the case of $N_{*, {\rm SN}}=0$, we assume that all mass loss results from stellar winds. In the case of $N_{*, {\rm SN}}>0$, we assume that all mass loss results from supernova explosions, and we model the corresponding kinetic and thermal energy injection due to the expanding blast-wave. At our mass resolution of $859~{\rm M}_\odot$ per gas cell, the energy-conserving/momentum-generating phase of supernova blast-wave expansion is unresolved, and so we follow the prescription introduced by~\cite{KimmCen14}: we explicitly inject the terminal momentum of the blast-wave into the set of gas cells $k$ that share faces with the nearest-neighbour cell to the star particle. We use the unclustered parametrisation of the terminal momentum derived from the high-resolution simulations of~\cite{Gentry17}, which is given by
\begin{equation}
\frac{p_{{\rm t}, k}}{{\rm M}_\odot~{\rm km~s}^{-1}} = 4.249 \times 10^5 N_{*, {\rm SN}} \Big(\frac{n_k}{{\rm cm}^{-3}}\Big)^{-0.06},
\end{equation}
with an upper limit imposed by the condition of kinetic energy conservation, as the shell sweeps through the gas cells $k$. The momentum is distributed among the facing cells as described in~\cite{2021MNRAS.505.3470J}.

The pre-supernova feedback from HII regions is implemented according to the model of~\cite{2021MNRAS.505.3470J}. This model accounts for the momentum injected by both radiation pressure and the `rocket effect', whereby warm ionised gas is ejected from cold molecular clouds, following the analytic work of~\cite{Matzner02} and \citet{KrumholzMatzner09}. Momentum is injected for groups of star particles with overlapping ionisation-front radii, which are identified using a Friends-of-Friends grouping prescription. The momentum is received by the gas cell closest to the luminosity-weighted centre of the Friends-of-Friends group, and is distributed to the set of adjoining neighbour cells. The gas cells inside the Str\"{o}mgren radii of each Friends-of-Friends group are heated to a temperature of $7000$~K, and are held above this temperature floor for as long as they receive ionising photons from the group. In contrast to~\cite{2021MNRAS.505.3470J}, we also explicitly and fully ionise the gas inside these Str\"{o}mgren radii, rather than relying on the chemical network to do so.

\section{Properties of the simulated galaxy} \label{Sec::sim-props}
Here we compute the key observable properties of our simulated galaxy and its interstellar medium, and demonstrate that their values are close to those obtained via direct observations in comparable galactic environments.

\subsection{Gas morphology and global interstellar medium structure} \label{Sec::discussion::observations}
Figure~\ref{Fig::morphology} shows the spatial distribution of the total (left), atomic (centre-left), total molecular (centre-right) and CO-luminous molecular (right) gas reservoirs at face-on and edge-on viewing angles, at a simulation time of $800$~Myr. The atomic and total molecular gas abundances are computed within our simplified chemical network during simulation run-time, as described in Section~\ref{Sec::sims}. To partition the molecular gas into its CO-luminous and CO-dark components, we post-process our simulation outputs using the {\sc Despotic} astrochemistry and radiative transfer model \citep{Krumholz13a, Krumholz14}, as described in Appendix~\ref{App::chem-postproc}, and delineate CO-dark from CO-bright material using a threshold that we detail below.

The qualitative structure of the CO-bright interstellar medium is similar to that observed by~\cite{Faesi18} in the nearby dwarf spiral galaxy NGC300, displayed in Figure 1 of~\cite{Kruijssen2019}. In Figure~\ref{Fig::tuning-fork}, we quantify the agreement between the structure of our simulated interstellar medium and the observed interstellar medium in NGC300, showing that the simulation reproduces the observed spatial decorrelation between regions of recent star formation (traced by young stars) and regions of high molecular gas surface density (traced by CO emission). To this end, we use the so-called ``tuning fork diagram'' that quantifies the relative bias of depletion time measured in the apertures centered either on dense gas or young stars \citep[e.g.,][]{Schruba10,Kruijssen2019}. The relative excess of dense gas or young stars in apertures of a given size ($x$-axis) results in the upper (blue) and lower (green) branches, respectively. The values calculated from observations by~\cite{Kruijssen2019} are given by the thin lines and open data points.

We calculate the dense gas and young stellar branches for our simulation as outlined in \citet{2021ApJ...918...13S}. In particular, we use molecular gas with a surface density of $\Sigma_{\rm H_2, CO} > \Sigma_{\rm H_2,CO,min} = 13\; M_\odot\;{\rm pc^{-2}}$ and a line-of-sight velocity dispersion of $< 1\;{\rm km\;s^{-1}} \times \Sigma_{\rm H_2, CO}/\Sigma_{\rm H_2,CO,min}$ as a proxy for the dense CO-bright gas; here $\Sigma_\mathrm{H_2,CO}$ is the H$_2$ surface density that would be inferred based on the computed CO luminosity, rather than the true H$_2$ surface density. We define this quantity precisely in Appendix \ref{App::chem-postproc}. The line-of-sight velocity dispersion is computed as the CO-weighted mean velocity dispersion of the Voronoi gas cells along rays perpendicular to the galactic mid-plane. We assume that the H$\alpha$ signal in young star-forming regions is produced by star particles with ages 2-5~Myr. The shaded regions show the $1\sigma$ and $2\sigma$ spread of values for our simulated interstellar medium between simulation times of $500$ and $800$~Myr.

\begin{figure}
	\includegraphics[width=\linewidth]{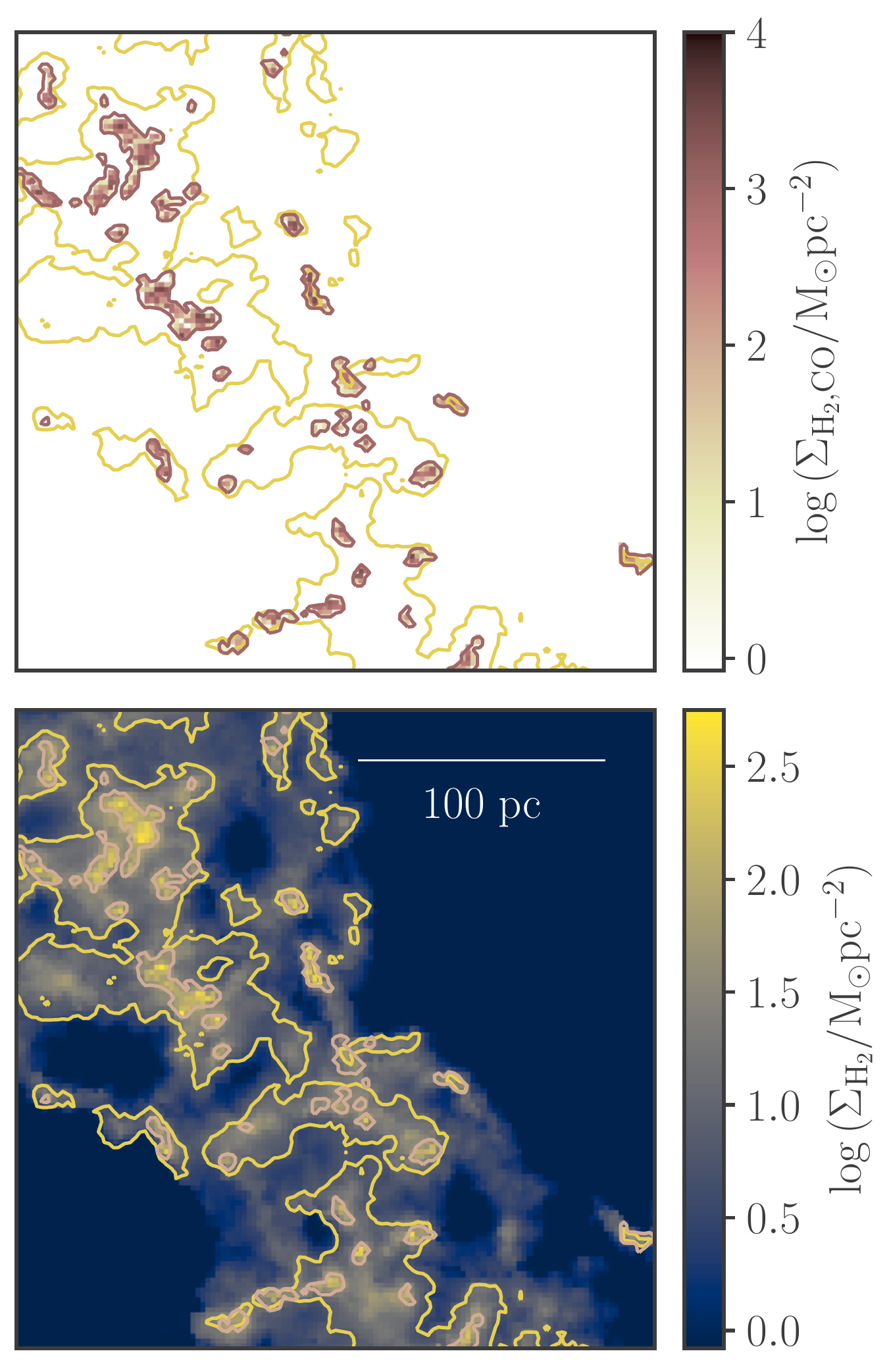}
	\caption{Thresholds for the identification of CO-luminous giant molecular clouds (pink isocontours) and their CO-dark envelopes down to a fractional projected abundance of $0.3$ (yellow isocontours). The upper panel shows the CO-luminous molecular hydrogen column density (pink) and the lower panel shows the total molecular hydrogen column density (blue-yellow).}
	\label{Fig::choosing-cloud-threshold.png}
\end{figure}

\begin{figure*}
	\includegraphics[width=\linewidth]{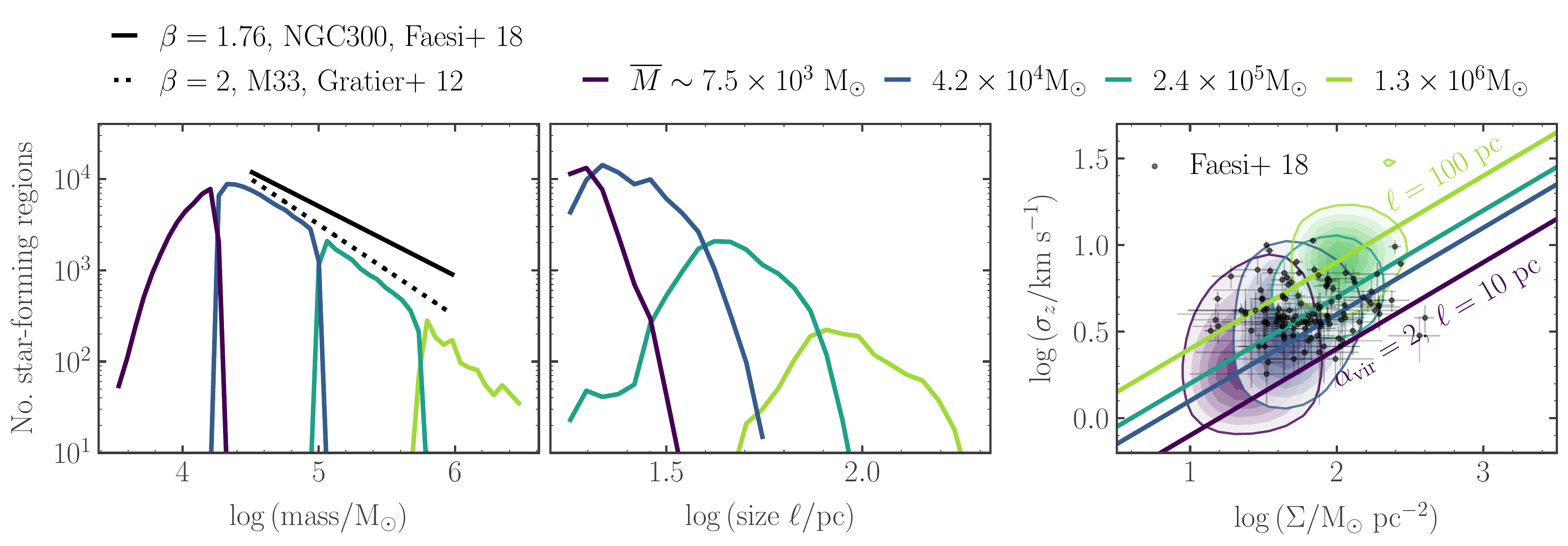}
	\caption{\textit{Left:} Mass distribution of CO-bright molecular clouds in our simulation, divided into four mass bins. The solid black line gives the observed powerlaw slope $\dd N/\dd M \propto M^{-\beta}, \beta = 1.76 \pm 0.07$ found by~\protect\cite{Faesi18} in NGC300. The dotted black line gives the powerlaw slope $\beta = 2 \pm 0.1$ found by~\protect\cite{2012A&A...542A.108G} in M33. \textit{Centre:} Size distribution of CO-bright molecular clouds in our simulation, divided into the same four mass bins. \textit{Right:} Molecular gas line-of-sight velocity dispersion as a function of the molecular gas surface density for the CO-luminous molecular clouds. The solid lines indicate virial parameters of $\alpha_{\rm vir}=2$ for spherical beam-filling clouds at the mean region size for each mass bin. Black data points and errorbars represent the resolved GMC sample of~\protect\cite{Faesi18} in NGC300.}
	\label{Fig::cloud-diagnostics}
\end{figure*}

The convergence, on large scales, of the upper and lower branches in Figure~\ref{Fig::tuning-fork} is a general property of star-forming galaxies~\citep[e.g.][]{2022MNRAS.516.3006K}, indicating the tight correlation between dense gas and young stars that is manifested in the near-linear relation between the molecular gas density and the star formation density observed in normal (non-starburst) star-forming galaxies on $>$kiloparsec scales \citep[e.g.,][]{WongBlitz2002,Bigiel08,Leroy+13}. On small scales, the branches diverge, as dense, CO-emitting gas and ionised, H$\alpha$-emitting gas become spatially decorrelated.

The precise shape of the branches in Figure~\ref{Fig::tuning-fork} is sensitive to the modelling of star formation and stellar feedback used in the numerical simulation~\citep{Fujimoto19, 2021ApJ...918...13S}. The agreement of this shape with the relation computed from observational data by~\cite{Kruijssen2019} indicates that our models produce a reasonable interstellar medium structure for galactocentric radii of $R = 2\text{--}6$~kpc. The inner 2~kpc of the simulation displays a stronger correlation (narrower branch opening) than is computed by~\cite{Kruijssen2019}. Although this is likely due to the existence of several high-mass molecular clouds in this region, which are removed (`masked') in their analysis, we exercise caution and exclude the region $R<2~{\rm kpc}$ from our subsequent analysis, as we cannot verify its agreement with the observational data.

For the remaining galactocentric radii $R>2~{\rm kpc}$, we demonstrate in Figure~\ref{Fig::observables} that our simulated dwarf spiral galaxy is similar in its atomic and molecular depletion times (upper panel) and its gas and stellar surface densities (lower panel) to NGC300. The simulated values are given by the bold lines, while the corresponding observed values from~\cite{2011MNRAS.410.2217W} and~\cite{Kruijssen2019} are given by the thin lines.

\begin{figure}
	\includegraphics[width=\linewidth]{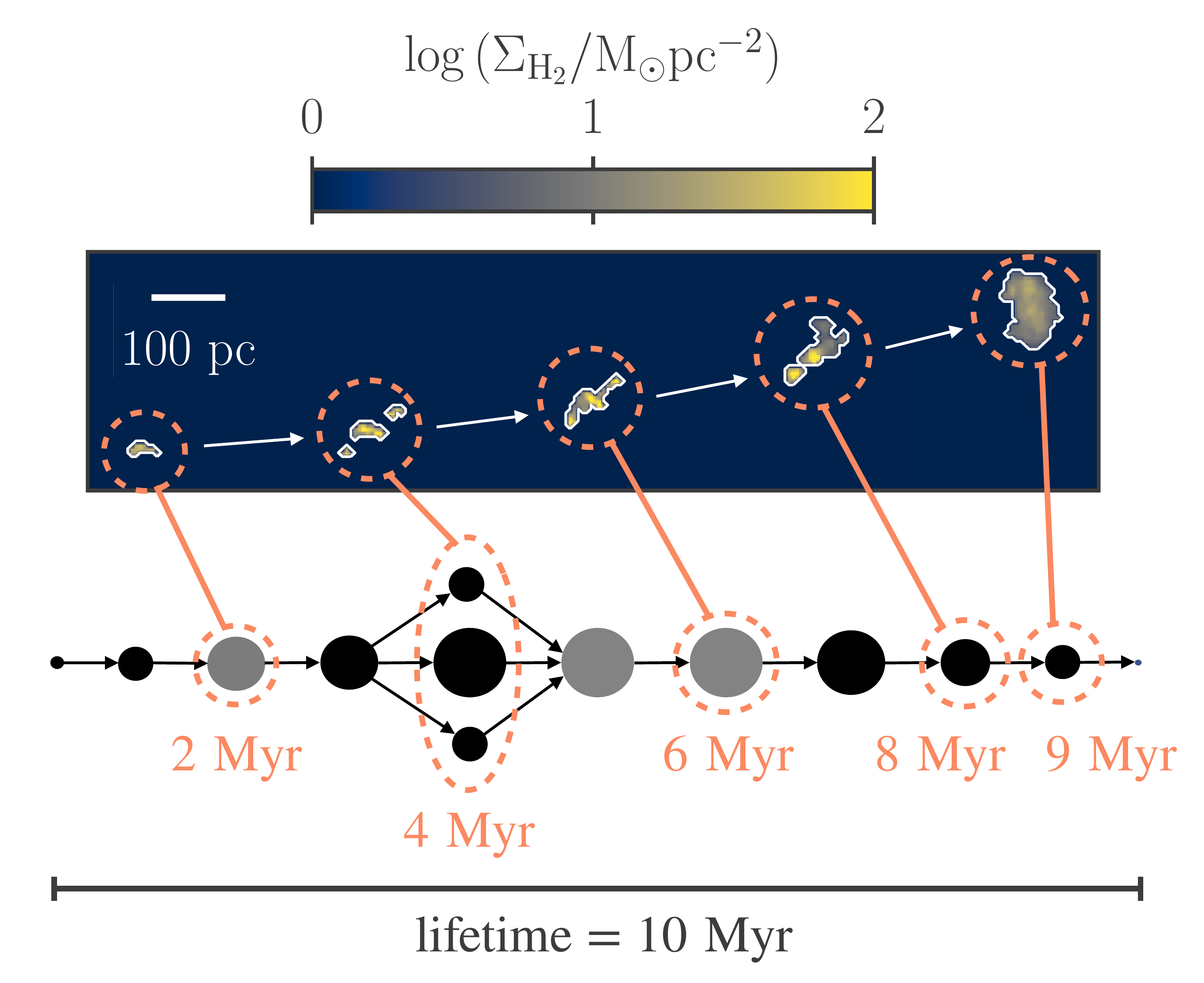}
	\caption{The lifecycle of a single simulated molecular cloud with a lifetime of $10$~Myr. The actual evolution of the ${\rm H_2}$ surface density as it moves across the galaxy is shown in the upper panel. Both the CO-bright and CO-dark ${\rm H_2}$ are shown. This region undergoes a split and re-merger $4$~Myr after its birth. The corresponding section of the cloud evolution network is shown in the lower half of the figure.}
	\label{Fig::network-fig}
\end{figure}

\subsection{The CO-bright molecular cloud population}
\subsubsection{Identification} \label{Sec::identification}
In this work, we identify CO-bright, observable clouds using isocontours of value $\log{(\Sigma_{\rm H_2, CO}/{\rm M}_\odot{\rm pc}^{-2})} = -1.5$ on the surface density $\Sigma_{\rm H_2, CO}$ of CO-bright molecular hydrogen perpendicular to the galactic mid-plane, shown in dark pink in Figure~\ref{Fig::choosing-cloud-threshold.png}.  This threshold corresponds to the natural break in the distribution of $\Sigma_{\rm H_2, CO}$ produced by our chemical post-processing, which is described in detail in Appendix~\ref{App::chem-postproc}. At surface densities higher than the threshold, gas cells contain at least some shielded, CO-dominated gas. At lower densities, CO exists only as a uniformly-mixed, unshielded, low-abundance component.

The yellow contours in Figure~\ref{Fig::choosing-cloud-threshold.png} enclose regions with a projected ${\rm H_2}$ abundance of $\Sigma_{\rm H_2}/\Sigma_{\rm gas} > 0.3$, accounting for 70~per~cent of the total ${\rm H_2}$ in the simulation, including CO-dark ${\rm H_2}$. We find that the majority of star formation ($78$~per~cent) occurs in the CO-luminous ${\rm H_2}$, despite the fact that this gas reservoir accounts for only $26$~per~cent of the galactic molecular hydrogen mass. In what follows, we will therefore focus on the CO-luminous giant molecular clouds. We note that 99~per~cent of the gas tracer particles occupying the CO-luminous state inside the pink contours of Figure~\ref{Fig::choosing-cloud-threshold.png} have previously resided in the CO-dark, ${\rm H_2}$-rich state denoted by the yellow contours. That is, the gas in our simulation almost always enters CO-luminous clouds by way of the CO-dark envelope.

\begin{figure}
	\includegraphics[width=\linewidth]{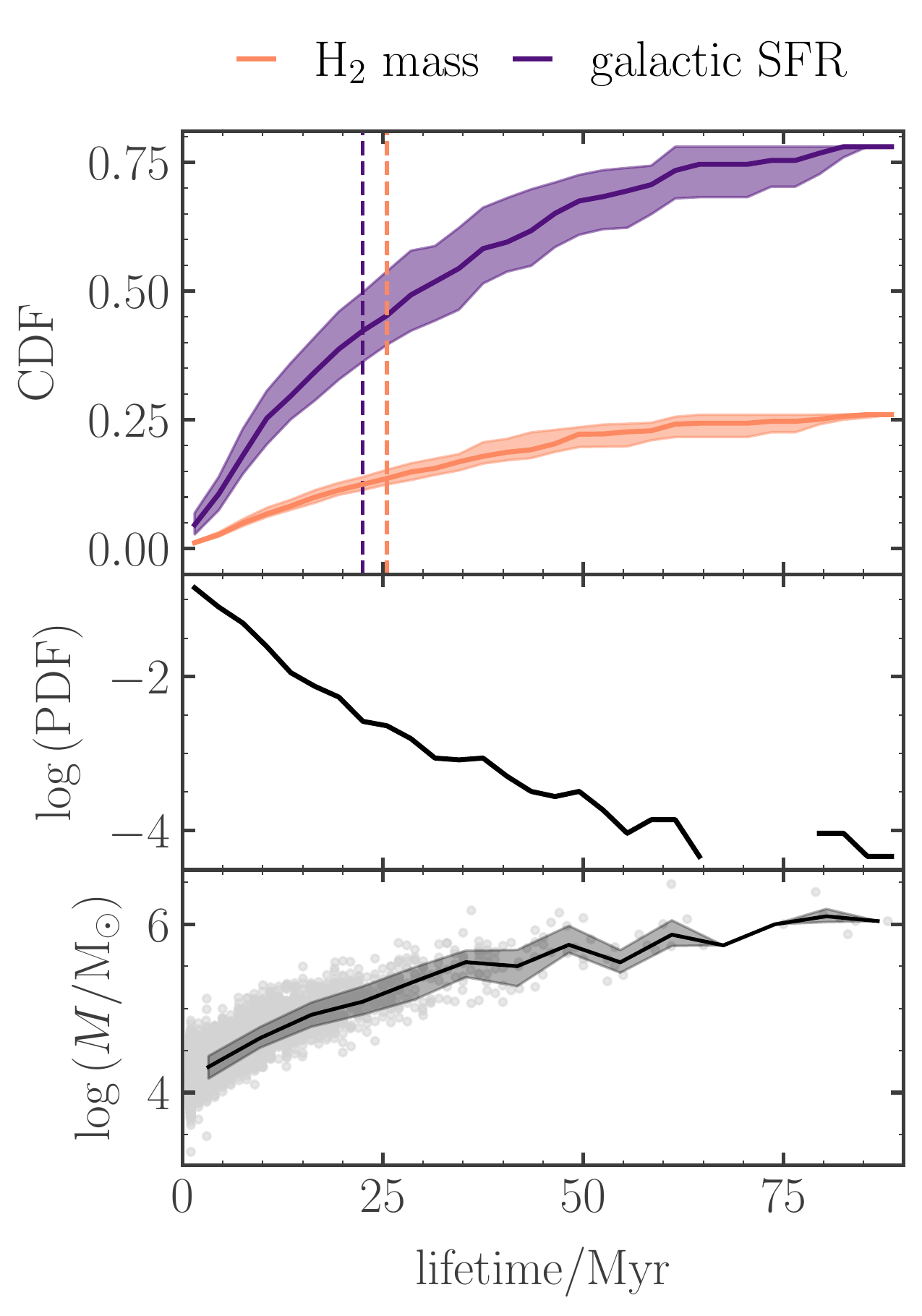}
	\caption{\textit{Upper panel:} Time-averaged cumulative fraction of the total galactic molecular mass (orange) and star formation rate (SFR; purple) accounted for by CO-bright clouds, as a function of their lifetime. The solid lines denote the time-averaged median values, while the shaded regions denote the corresponding interquartile ranges. The normalisation of the CDFs reflects the fact that CO-dark gas accounts for 22~per~cent of star formation and 74~per~cent of galactic ${\rm H_2}$ (see Section~\protect\ref{Sec::identification}).  The orange and purple dashed vertical lines represent the median cloud lifetimes, weighted by the ${\rm H_2}$ mass and by galactic SFR, respectively. \textit{Centre panel:} Probability distribution of the molecular cloud lifetime by number. \textit{Lower panel:} Median and interquartile range of the molecular cloud peak mass, as a function of the cloud lifetime. Grey circles represent the values for individual GMCs.}
	\label{Fig::intro-CDFs}
\end{figure}

\subsubsection{Observational checks}
An important check for the applicability of our simulation to the real Universe is whether the simulated population of CO-bright molecular clouds has similar properties to the observed population in a dwarf spiral galaxy. In Figure~\ref{Fig::cloud-diagnostics}, we compare the instantaneous mass distribution (left panel) and size distribution (centre panel) of our simulated molecular clouds across all simulation times to the population of resolved clouds in the inner disc ($R \la 3$~kpc) of NGC300~\citep{Faesi18} and in the outer regions ($R \ga 2$~kpc) of M33~\citep{2012A&A...542A.108G}. We divide the simulated cloud population into four mass bins, to highlight the correspondence between the cloud mass, surface density and velocity dispersion.

We see that the slope of the simulated cloud mass distribution agrees well with the observed slope of $\beta = 2 \pm 0.1$ in the outer regions of M33, but is significantly steeper than the corresponding slope in NGC300. It does not display a truncation at high masses, as seen for NGC300 (not shown in the figure). Given that our simulated cloud population is made up of clouds at $R>2~{\rm kpc}$, we would expect that our results are more closely-comparable to the M33 sample. Note however that such a comparison should be interpreted with caution, given the sensitivity of the mass function slope to the method of cloud identification used~\citep{2009ApJ...699L.134P,Hughes13b,2021MNRAS.502.1218R}. Nevertheless, we note that the distribution of masses for the molecular clouds in our simulation is in reasonable agreement with observed values.

Comparing the coloured contours to the coloured lines in the right-hand panel, we see that the cloud population generally follows a line of constant virial parameter, as seen in the observations of resolved clouds in NGC300~\citep[black data points,][]{Faesi18}. Higher-mass clouds are closer to virial equilibrium on average, with higher levels of gravitational boundedness.

\begin{figure}
  \includegraphics[width=\linewidth]{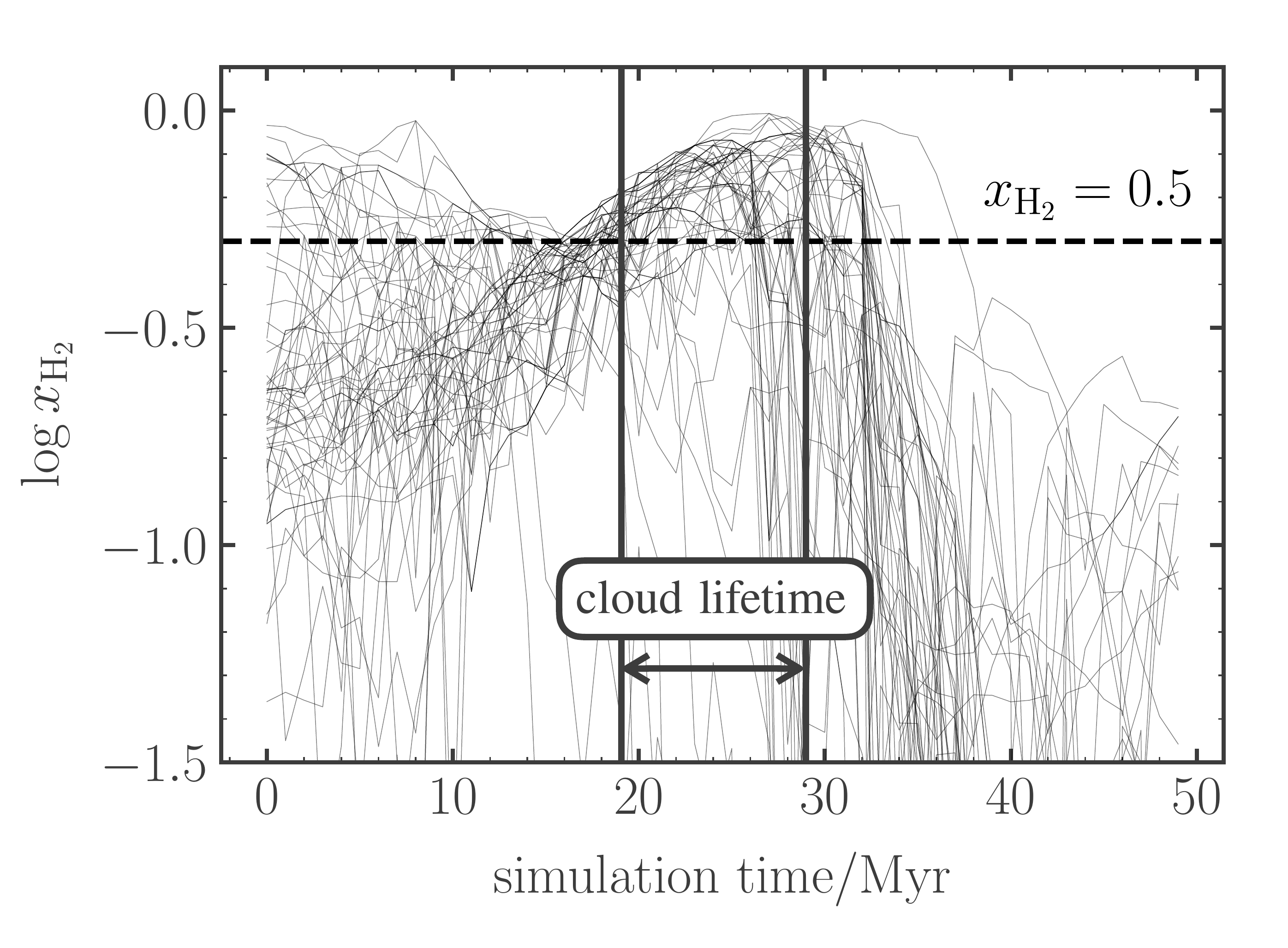}
  \caption{The evolution of the molecular hydrogen abundance $x_{\rm H_2}$ of individual Lagrangian gas parcels (thin black lines), as they transit through a CO-bright molecular cloud of lifetime $10$~Myr. This is the same molecular cloud shown in Figure~\protect\ref{Fig::network-fig}. The black vertical lines denote the times at which the cloud appears and disappears, based on our cloud-tracking algorithm.}
  \label{Fig::network-fig-2}
\end{figure}

\subsubsection{Tracking and evolution}
We track the temporal evolution of the simulated molecular cloud population via the algorithm described in~\cite{2021MNRAS.505.1678J}. In brief, the position of the two-dimensional isocontour enclosing a star-forming region at simulation time $t$ is projected forward by the time-step $\Delta t = 1~{\rm Myr}$ of our simulation output, using the velocities of the gas cells in the region. A pair of star-forming regions is temporally linked as parent and child if there exists any overlap between the projected contour of the parent at time $t$ and the contour outlining the child at time $t+\Delta t$.

Via this tracking procedure, we produce the \textit{cloud evolution network} for the entire simulation, composed of $\sim 8000$ complete, independent segments between simulation times of $t=500$ and $800$~Myr, and between galactocentric radii of $R=2$ and $6$~kpc. These segments correspond to time-evolving molecular clouds, which we use in the analysis presented in Section~\ref{Sec::results}. The lifetimes of these clouds are calculated as the end-to-end time between the formation of the first parent cloud and the destruction of the last child, accounting for all cloud mergers and splits, as shown in Figure~\ref{Fig::network-fig}.

\begin{figure*}
	\includegraphics[width=\linewidth]{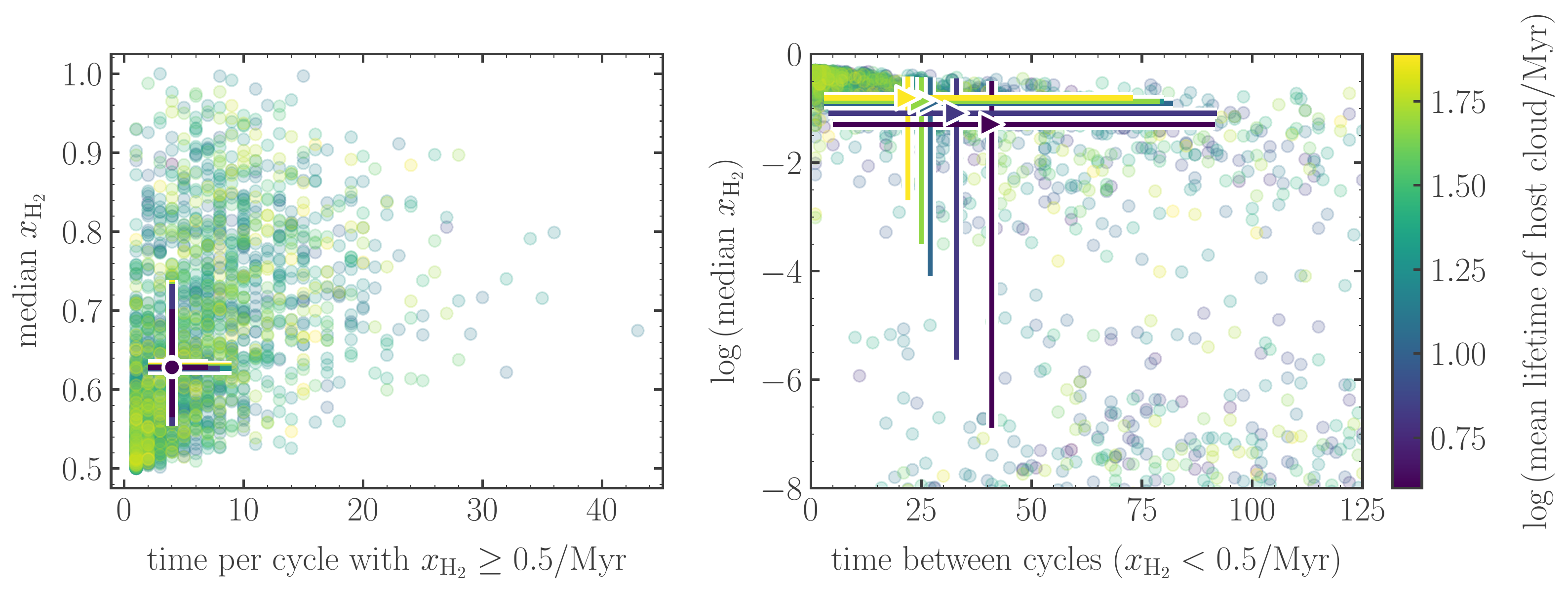}
	\caption{The time spent by gas tracer particles in the ${\rm H_2}$-rich ($x_{\rm H_2} \geq 0.5$; left) and ${\rm H_2}$-poor ($x_{\rm H_2} < 0.5$; right) phases, for tracer particles that transit through CO-bright molecular clouds. The transparent data points give the values for $1/5000$th of the cycles undergone by tracer particles, while the solid data points and lines give the median values and interquartile ranges along each axis, computed in bins of molecular cloud lifetime as indicated by the colours of the solid points. All points are coloured according to the lifetime of the cloud through which the tracer particle transits. The time-scales spent in the ${\rm H_2}$-poor state are strictly lower limits, as $60$~per~cent of the values abut the beginning or end of the simulation window at $t=500$ or $t=800$~Myr, and so are truncated.}
	\label{Fig::moleculartime-vs-meanIH2-LCOGMCs}
\end{figure*}

\section{Results} \label{Sec::results}
In this section, we combine our analysis of the molecular cloud network with results derived from the tracer particles in our simulation to study the formation and destruction of ${\rm H_2}$ molecules within CO-bright molecular clouds, as a function of the cloud mass and cloud lifetime. We begin with the distribution of masses and lifetimes derived from the tracking of Eulerian clouds in Section~\ref{Sec::Eulerian-results}, turn to the results from the tracer particles in Section~\ref{Sec::xH2-vs-lifetime}, and combine the two viewpoints in Section~\ref{Sec::competition-model}. In Section~\ref{Sec::clustering}, we discuss the impact of the Eulerian cloud mass distribution on the clustering of supernovae.

\subsection{
The Eulerian view: long-lived clouds}
\label{Sec::Eulerian-results}
In the upper panel of Figure~\ref{Fig::intro-CDFs}, we show the fraction of the simulated galactic ${\rm H_2}$ reservoir (orange) and star formation rate (purple) that is accounted for by CO-bright molecular clouds of different lifetimes. We see that half of the star formation occurring in CO-bright molecular clouds (40~per~cent of the galactic total) is accounted for by molecular clouds that live longer than $25$~Myr, despite the fact that the most-common lifetime for molecular clouds, by number, is $<10~{\rm Myr}$ (centre panel of Figure~\ref{Fig::intro-CDFs}). Clouds with lifetimes $>25$ Myr also contain about half the total CO-luminous H$_2$ mass in the galaxy.

In the lower panel of Figure~\ref{Fig::intro-CDFs}, we demonstrate that molecular cloud lifetime is tightly correlated and monotonically increasing with the peak cloud mass achieved throughout this lifetime. We will discuss this proportionality in detail in Section~\ref{Sec::competition-model}, but for now we simply note that `long-lived' molecular clouds are synonymous with `massive' clouds. Therefore, a substantial fraction of the CO-bright molecular gas in our dwarf spiral galaxy simulation is contained in clouds that are both massive and long-lived.

\subsection{
The Lagrangian view: short-lived H$_2$ molecules
}
\label{Sec::xH2-vs-lifetime}
We now examine the chemical survival time of molecular hydrogen in CO-bright molecular clouds, by tracking the ${\rm H_2}$ abundance of Lagrangian gas parcels (tracer particles) as they transit through these clouds. We consider a gas parcel to have passed through a cloud if it moves within the isocontour defining its edge at any time. We show the time evolution of the H$_2$ mass fraction $x_\mathrm{H_2} = \rho_\mathrm{H_2}/\rho_\mathrm{gas}$ for the collection of gas parcels transiting through a single example cloud in Figure~\ref{Fig::network-fig-2}. We omit tracer particles that remain at molecular hydrogen abundances below $x_{\rm H_2}=10^{-5}$ throughout the simulation run-time from our analysis; this cut excludes those tracer particles that are far from the galactic mid-plane but appear inside the isocontour in projection.

As the plot shows, individual fluid elements experience a wide range of variations in $x_\mathrm{H_2}$, with many experiencing rapid fluctuations in the H$_2$ abundance, particularly as the cloud is disrupted. To quantify this behaviour, for each tracer particle that passes through a CO-bright molecular cloud, we identify all the contiguous time periods for which it has $x_\mathrm{H_2} < 0.5$ and $ > 0.5$; for each such period, we record its duration, the median (in time) value of $x_\mathrm{H_2}$ during it, and the lifetime of the CO-bright cloud through which it passed.\footnote{For particles that have multiple episodes of $x_\mathrm{H_2}>0.5$ with a period of $x_\mathrm{H_2}<0.5$ between them, we take the molecular cloud lifetime for the H$_2$-poor phase to be the lifetime of the molecular cloud through which the particle passed during the preceding H$_2$-rich phase. Essentially no tracer particles pass through two different CO-bright molecular clouds during a single cycle of having $x_\mathrm{H_2} > 0.5$ and $<0.5$.}
In the left and right panels of Figure~\ref{Fig::moleculartime-vs-meanIH2-LCOGMCs}, we plot the median $x_\mathrm{H_2}$ versus duration, with points colour-coded by molecular cloud lifetimes; the left panel shows intervals of $x_\mathrm{H_2} > 0.5$, while the right shows $x_\mathrm{H_2} < 0.5$. The plot shows that gas rapidly cycles in and out of an ${\rm H_2}$-dominated state ($x_{\rm H_2}>0.5$), in qualitative agreement with~\cite{Semenov17}. The time spent by gas parcels in the ${\rm H_2}$-rich state is short, ranging from $1$ to $\sim 20$~Myr with a median value of $4$~Myr. The median abundance of molecular hydrogen over the duration of each cycle is $0.63$. Crucially, the distributions of ${\rm H_2}$ molecule lifetimes and $x_{\rm H_2}$ values are almost identical between the samples of molecular clouds with different lifetimes (different colours). Thus we conclude that \textbf{independent of the lifetime of the host molecular cloud, the chemical survival time of of ${\rm H_2}$ molecules is about 4~Myr.}

\begin{figure*}
  \includegraphics[width=\linewidth]{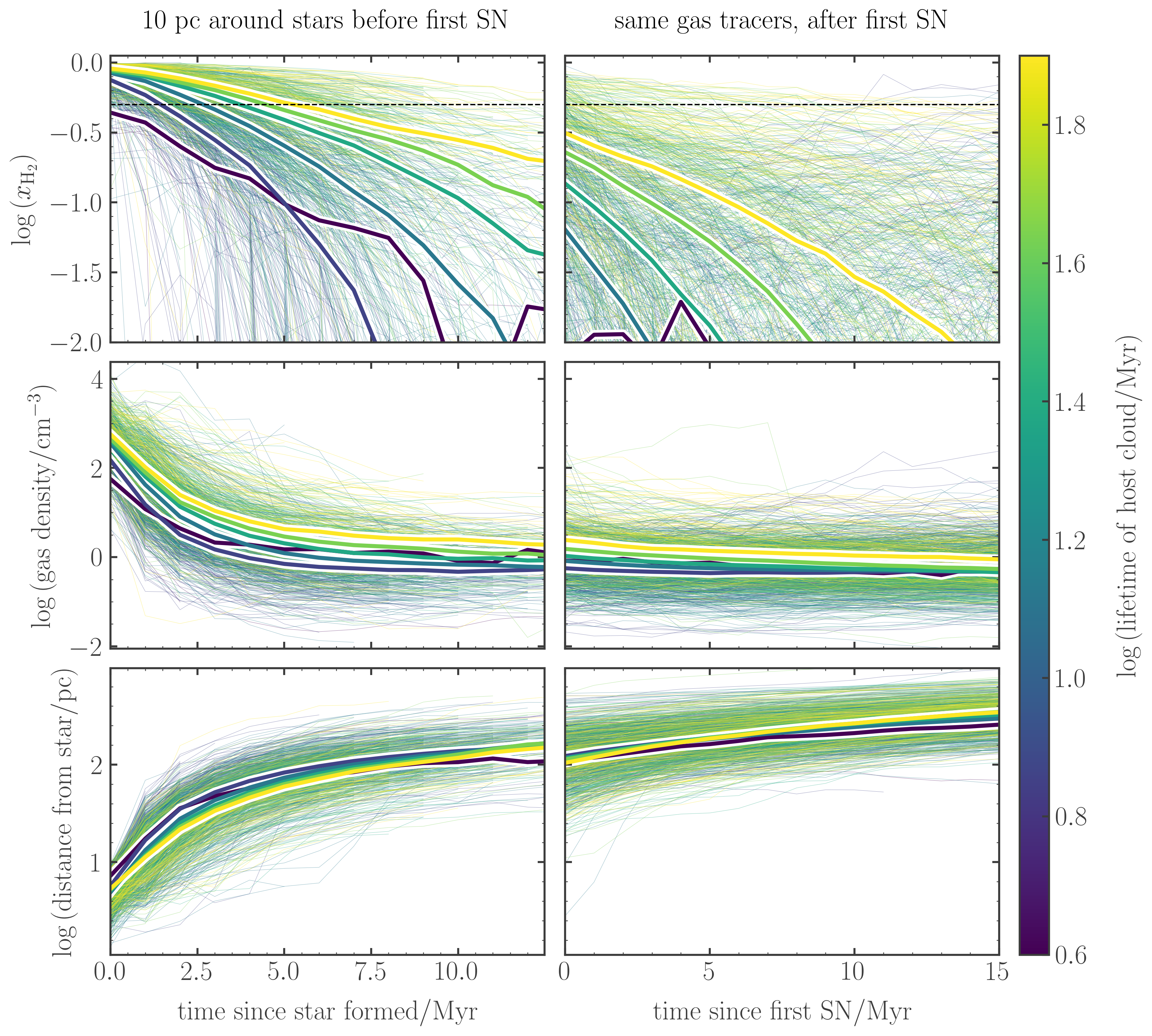}
  \caption{The behaviour of gas parcels (tracer particles) around all young star particles in our simulation. The colorbar corresponds to the lifetime of the host molecular cloud. Thin lines show the median values for the gas tracer particles that are located within 10 pc of a star particle at the instant of its formation, while thick lines show the medians across this population, computed in bins of molecular cloud lifetime. The left column shows the time evolution starting at the instant of star formation, while the right column shows the time evolution starting from the moment when the star particle produces its first supernova. We see that early, pre-supernova feedback (left-hand column) ionises and drives the gas away from young stars on time-scales between $2$ and $5$~Myr across all molecular clouds, with the ejection being slightly slower in the longer-lived, higher-mass regions. Supernovae (right-hand column) generally occur when the gas has already been expelled from an ${\rm H_2}$-rich state.}
  \label{Fig::FB-IH2-vs-time}
\end{figure*}

\begin{figure}
	\includegraphics[width=\linewidth]{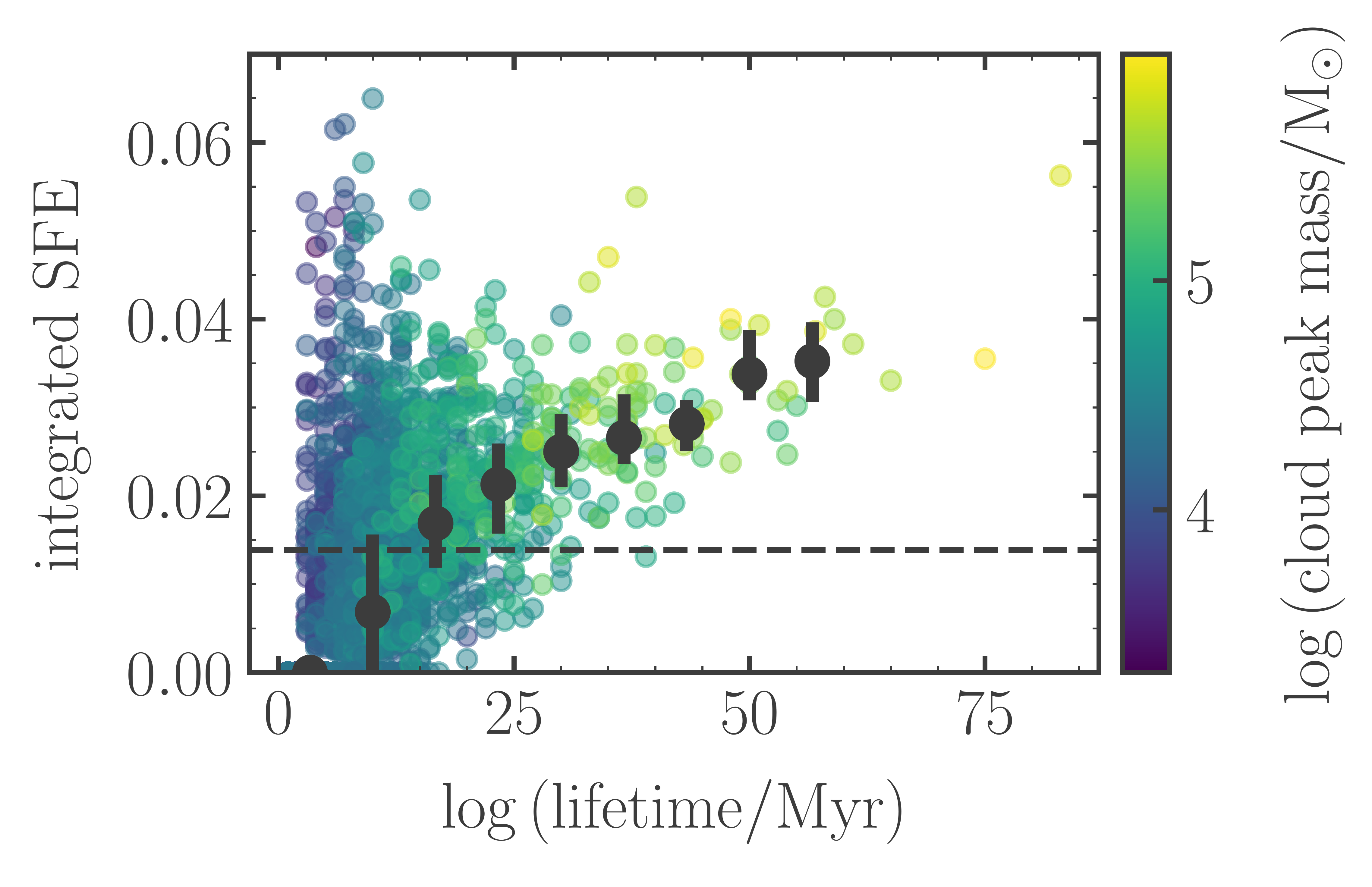}
	\caption{Integrated star formation efficiency over the lifetime of a CO-bright molecular cloud, as a function of its lifetime. The black data points and error bars correspond to the median values and interquartile ranges in each lifetime interval. The horizontal dashed line gives the total stellar mass produced in all clouds, divided by the total mass of all clouds, which is equal to the total SFE within the simulation. The transparent data points correspond to the values for individual molecular clouds, coloured by the peak mass they attain during their lifetimes.}
	\label{Fig::intSFE}
\end{figure}

The right-hand panel of Figure~\ref{Fig::moleculartime-vs-meanIH2-LCOGMCs} shows that the time spent by gas parcels in the ${\rm H_2}$-poor state ($x_{\rm H_2}<0.5$) \textit{is} correlated with the lifetime of the host molecular cloud, albeit weakly. Gas parcels ejected from higher-mass, longer-lived clouds (yellow) return to the ${\rm H_2}$-dominated state after shorter periods of time, and retain higher median ${\rm H_2}$ abundances, than those associated with shorter-lived, lower-mass clouds (purple). That is, ${\rm H_2}$ is not as efficiently or quickly dissociated in higher-mass molecular clouds. Gas parcels ejected from these clouds are more likely to become `trapped', and cycle back to an ${\rm H_2}$-dominated state multiple times: the typical number of cycles for tracer particles transiting through clouds of lifetime $<25~{\rm Myr}$ is $1$, but for regions of lifetime $>25~{\rm Myr}$, the number of cycles is between $2$ and $4$.

In Figure~\ref{Fig::FB-IH2-vs-time}, we demonstrate that the ejection of gas from the ${\rm H_2}$-dominated state is driven by early stellar feedback. To construct this figure, we tag every Lagrangian tracer particle that lies within 10 pc of a star particle at the moment of its formation. For each tagged tracer, we also identify the lifetime of the GMC within which that tracer resides. On the left-hand side of Figure~\ref{Fig::FB-IH2-vs-time}, we plot the time evolution of the H$_2$ fraction, total gas density, and distance from the star for each of these tracers, terminating each line at the time of the star particle's first supernova. The thin lines in the figure correspond to the histories of individual tracers, and the thick lines to median values we obtain by binning the tracers by molecular cloud lifetime. Although the variation of individual tracks is large, we see that \textbf{on average gas is efficiently ejected from the ${\rm H_2}$-dominated state by pre-supernova feedback on time-scales between $2$ and $5$~Myr.} The gas is pushed out to a distance of $\sim 80$~pc and down to a density of between $1$ and $10~{\rm cm}^{-3}$ (close to the bulk mean density of hydrogen in the interstellar medium), where its ${\rm H_2}$ is dissociated by the interstellar radiation field. However, the efficiency of this ejection is a strong function of molecular cloud lifetime: gas parcels in higher-mass, longer-lived molecular clouds start at higher densities and H$_2$ fractions, and after star formation they retain higher density and ${\rm H_2}$ abundance out to larger distances from the newly-formed star particle.

On the right-hand side of Figure~\ref{Fig::FB-IH2-vs-time}, we show the further time evolution of these same Lagrangian tracers, but starting at the instant of the first supernova. The plot demonstrates that the role of supernova feedback in destroying ${\rm H_2}$ is secondary to that of pre-supernova feedback. By the time that stars explode as supernovae, the surrounding gas is already relatively ${\rm H_2}$-poor ($x_{\rm H_2}<0.5$) and close to the mean density of the bulk interstellar medium ($\sim 1~{\rm cm}^{-3}$). The main (partial) exception to this is is in the most massive and longest-lived clouds, where supernovae do further decrease the H$_2$ fraction. However, even for these clouds the mean H$_2$ fraction is below 50\% by the time the first supernovae occur.

The increased difficulty of destroying ${\rm H_2}$ molecules in higher-mass/long-lived molecular clouds by pre-supernova feedback results in an increase in the integrated star formation efficiency with cloud mass and lifetime. In Figure~\ref{Fig::intSFE}, we calculate the efficiency as the fraction of tracer particles that are transferred from gas to star particles as they transit through molecular clouds, and find an increase from $\sim 1$ to $\sim 4$~per~cent for an increase in cloud lifetime from $\sim 10$ to $\sim 90$~Myr. The horizontal dashed line represents the median integrated star formation efficiency for the entire CO-luminous gas reservoir. This corresponds to the star formation efficiency associated with the most common cloud lifetime of $\sim 10$~Myr, as reported in Figure~\ref{Fig::intro-CDFs}.

Interestingly, Figure~\ref{Fig::intSFE} shows that the lifetimes of molecular regions, $t_{\rm GMC}$, correlate near-linearly with their integrated SFE, $\epsilon_{\rm int}$. Such a correlation might explain the near-linear correlation between $\Sigma_{\rm H_2}$ and $\dot{\Sigma}_\star$ observed on $\gtrsim$kiloparsec scales in normal (non-starburst) galaxies. The depletion time of molecular gas can be expressed as $\tau_{\rm H_2} \equiv \Sigma_{\rm H_2}/\dot{\Sigma}_\star \sim t_{\rm GMC}/\epsilon_{\rm int}$, and therefore, if $\epsilon_{\rm int}$ scales only with $t_{\rm GMC}$ then $\tau_{\rm H_2}$ will be constant, i.e., $\Sigma_{\rm H_2} \propto \dot{\Sigma}_\star$ independent of kiloparsec-scale environment \citep[see][for more detail]{Semenov19}. Indeed, we find that in the region where we apply our analysis, $\tau_{\rm H_2}$ is close to constant (see Figure~\ref{Fig::observables}).

\begin{figure*}
\flushleft
    \includegraphics[width=.75\linewidth]{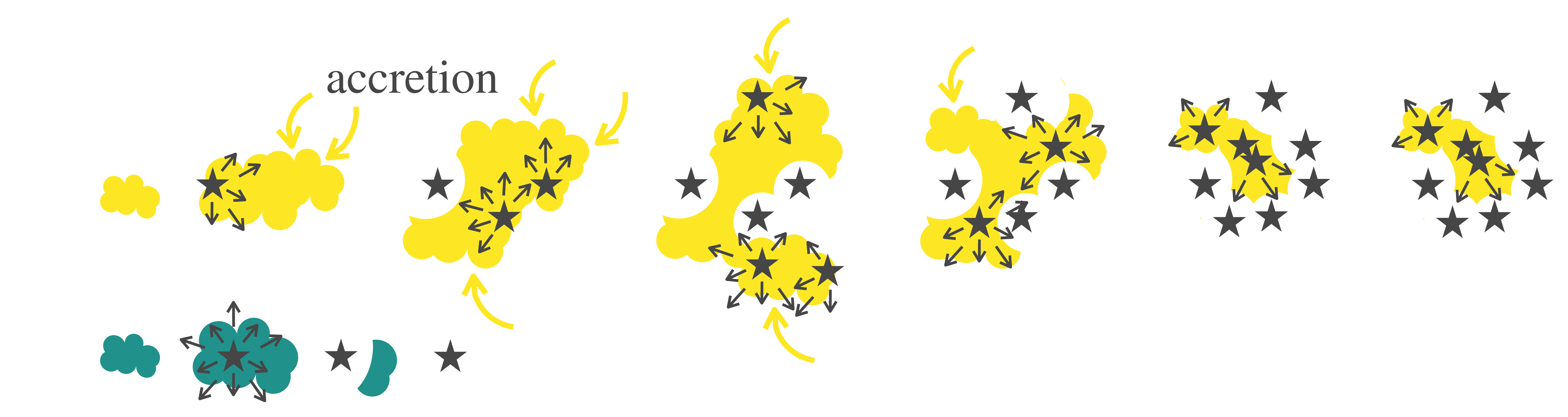}
	\includegraphics[width=0.9\linewidth]{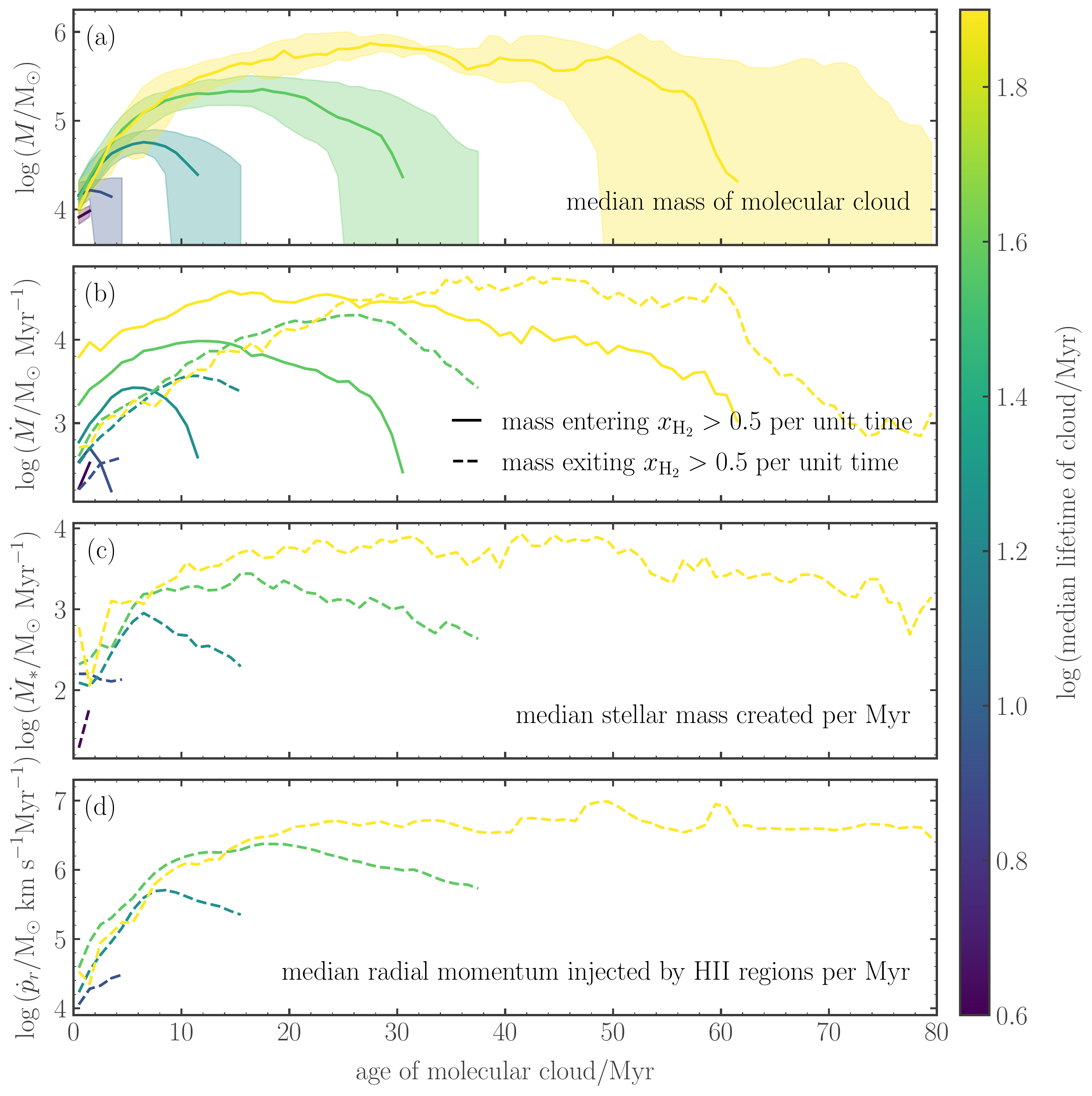}
	\caption{
        Cloud properties as a function of time since cloud formation, computed as medians over bins of total cloud lifetime, from the shortest-lived clouds (purple) to the longest-lived (yellow). Panel \textit{(a)} shows instantaneous cloud mass, with the solid lines showing the medians and the shaded bands showing the interquartile range. Panel \textit{(b)} shows mass flux into (solid) and out of (dashed) the H$_2$-dominated phase, $x_\mathrm{H_2} > 0.5$. Panel (c) shows star formation rate, and (d) shows rate of radial momentum injection by HII regions; both of these rates are computed using a 5 Myr averaging window to suppress fluctuations due to the stochasticity of star formation. Note that the total mass follows a symmetrical pattern of mass evolution from formation to destruction, and that the mass loss rate from the H$_2$-rich phase, star formation rate, and momentum injection rate are all strongly correlated. The schematic at the top shows a possible lifecycle for a short-lived (green) and a long-lived (yellow) molecular cloud, including the effects of accretion and feedback-induced gas ejection.}
	\label{Fig::accr-decr-vs-time}
\end{figure*}

\subsection{
Reconciling the views: molecular cloud evolution driven by the competition between accretion and ejection
} \label{Sec::competition-model}
We have shown in Section~\ref{Sec::Eulerian-results} that CO-bright molecular clouds are relatively long-lived. By contrast, in Section~\ref{Sec::xH2-vs-lifetime} we have shown that their constituent ${\rm H_2}$ molecules are very short-lived. In Figure~\ref{Fig::accr-decr-vs-time}, we show how these two results can be reconciled.

In panel $(a)$, we explicitly show the mass evolution of all molecular clouds in our simulation, as a function of time. To aid visualisation, we divide the clouds into five bins of lifetime, indicated by colour. The median values at each time are given by the solid lines and the shaded regions indicate the corresponding interquartile ranges. In agreement with the trend shown in Figure~\ref{Fig::intro-CDFs}, longer-lived molecular clouds achieve higher peak masses. The mass evolution of the molecular clouds in the simulation is remarkably symmetrical. All clouds begin and finish at the same mass, which is determined by the onset of effective ${\rm H_2}$ self- and dust-shielding at our mass resolution of $859~{\rm M}_\odot$. The clouds increase in mass for the first half of their lives, and decrease in mass for the second half.

In panel $(b)$ of Figure~\ref{Fig::accr-decr-vs-time}, we demonstrate that the mass evolution of the molecular clouds can be described by the competing accretion and ejection of gas into and out of a state of high ${\rm H_2}$ abundance. The solid lines show the median mass of Lagrangian gas parcels (tracer particles) entering the ${\rm H_2}$-dominated state per unit time per molecular cloud; the dashed lines show the total mass exiting this state per unit time. The masses of the molecular clouds (upper panel) switch from increasing to decreasing when the rate of mass ejection crosses above the rate of mass accretion. From mass conservation, the evolution of cloud masses can be described by the simple formula
\begin{equation}
M(t) = \int^t_0{\dd t^\prime \left[ \dot{M}_{\rm accr}(t^\prime) - \dot{M}_{\rm ej}(t^\prime) \right]}
,
\end{equation}
where $M(t)$ is the instantaneous mass of the cloud, and $\dot{M}_{\rm accr}$ and $\dot{M}_{\rm ej}$ are the instantaneous rates of ${\rm H_2}$ mass accretion and ejection, respectively. The total lifetime of the cloud is therefore given by
\begin{equation}
\int^{t_{\rm life}}_0 {\dd t^\prime \: \dot{M}_{\rm accr}(t^\prime)} = \int^{t_{\rm life}}_0 {\dd t^\prime \: \dot{M}_{\rm ej}(t^\prime)}.
\end{equation}
We note that this finding is broadly consistent with the analytic models proposed by \citet{Goldbaum11a}, \citet{2015A&A...580A..49I}, \citet{2017MmSAI..88..533B} and \citet{2017ApJ...836..175K}. These authors argue that molecular cloud lifetimes are determined primarily by external accretion rates, and that the galactic distributions of molecular cloud mass and star formation rate results from a competition between gas accretion driven by the compression of gas at the interfaces between large-scale converging flows and gas ejection driven by galactic shear and stellar feedback.

Panels $(c)$ and $(d)$ of Figure~\ref{Fig::accr-decr-vs-time} demonstrate that the rate of molecular gas ejection from clouds is correlated in time with the instantaneous star formation rate and, closely related, the momentum injection rate from HII regions. The median rate of star formation per cloud is indicated by the dashed lines in panel $(c)$, and the corresponding radial momentum injected by the HII regions is indicated by the dashed lines in panel $(d)$. The star formation rate tracks the cloud mass, and therefore the rate of molecular gas ejection by stellar feedback also tracks the cloud mass. It begins at the same value for all clouds, and diverges only after it crosses the mass accretion rate and so begins to destroy the cloud.

\begin{figure}
	\includegraphics[width=\linewidth]{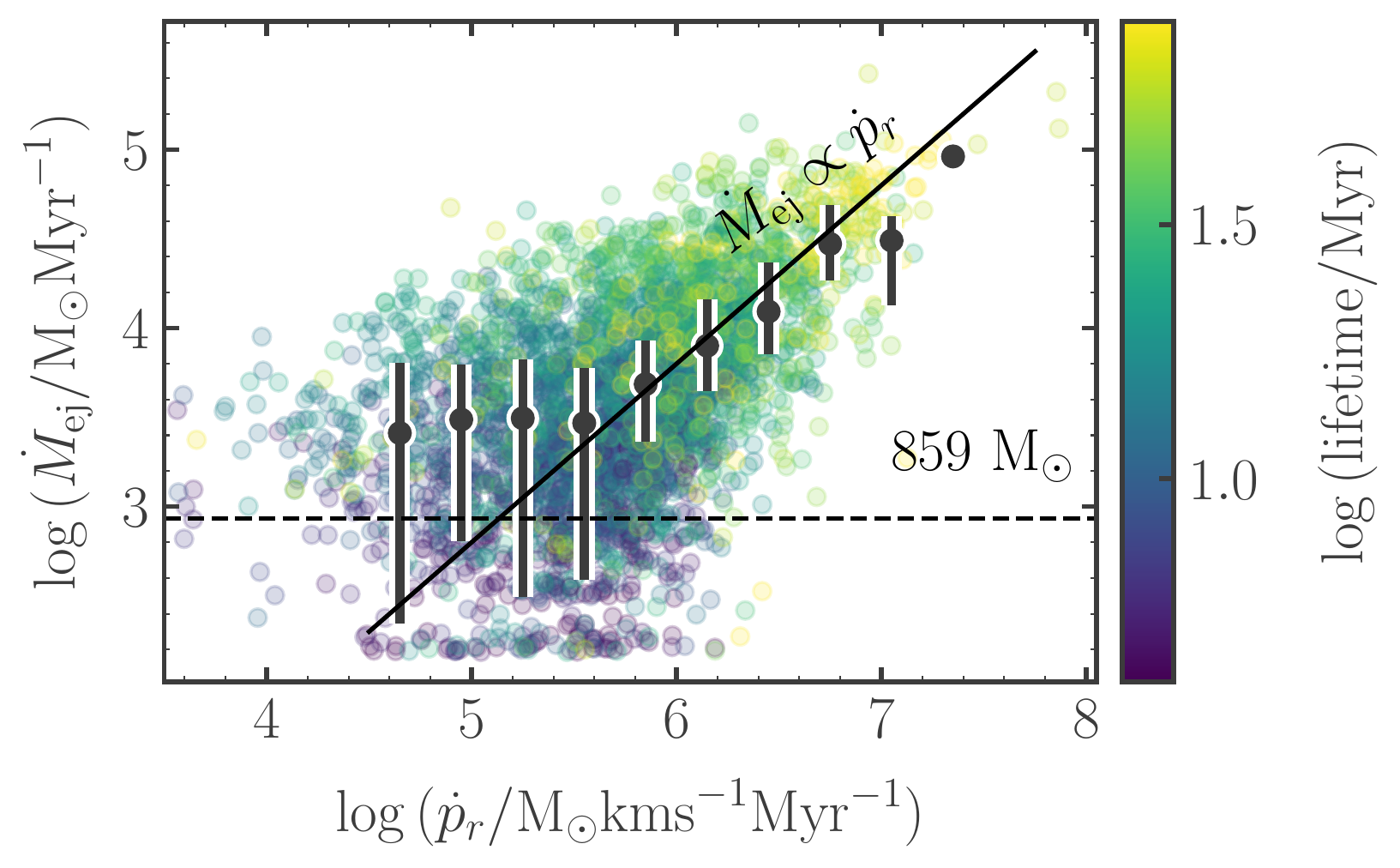}
	\caption{The rate of mass ejection $\dot{M}_{\rm ej}$ as a function of the radial momentum injection rate $\dot{p}_r$, for all molecular clouds in the simulation that are receiving feedback momentum. The scatter plot shows $1/5$th of the total measurements for the cloud population, coloured according to the lifetime of the cloud. The black points and errorbars give the median and interquartile range values in $10$ different mass bins. To guide the eye, the black solid line corresponds to the case of direct proportionality between mass ejection and momentum injection (not a fit). The black dashed horizontal line shows the resolution limit of the simulation, which corresponds to ejecting one resolution element of mass $\Delta M = 859$ M$_\odot$ over the time interval $\Delta t = 1$ Myr that we use in our numerical computation of $\dot{M}_\mathrm{ej}$.}
	\label{Fig::ejmass-vs-mom}
\end{figure}

We demonstrate the proportionality between the mass ejection rate $\dot{M}_{\rm ej}$ and the rate of momentum injection $\dot{p}_r$ by HII regions more explicitly in Figure~\ref{Fig::ejmass-vs-mom}. The fact that $\dot{M}_{\rm ej} \propto \dot{p}_r$ can be understood in terms of the momentum required to eject material from a (spherical) bounding surface at radius $r_{\rm shield}$ around each young star particle, where $r_{\rm shield}$ is the radius at which $x_{\rm H_2}$ drops to $0.5$, set by the extent of self- and dust-shielding. In this case,
\begin{equation}
    \dot{M}_{\rm ej} \sim \frac{\dot{p}_r}{v_{\rm esc}} \sim \frac{1}{r_{\rm shield}}\sqrt{\frac{3}{8\pi G \rho}} \dot{p}_r,
\end{equation}
where $v_\mathrm{esc}$ is the escape speed at $r_\mathrm{shield}$, which we write in terms of the mean volume density $\rho$ of the gas inside $r_{\rm shield}$. The variation in the values of $\rho$ and $r_{\rm shield}$ is not very large, resulting in the proportionality $\dot{M}_{\rm ej} \propto \dot{p}_r$ shown in  Figure~\ref{Fig::ejmass-vs-mom}.

\textbf{We therefore find that the rate of mass ejection $\dot{M}_{\rm ej}$ across all molecular clouds in our simulation can be straight-forwardly parametrised in terms of the rate of momentum injection $\dot{p}_r$ into these regions by early feedback.}

\subsection{The clustering of supernovae in massive, long-lived molecular clouds}
\label{Sec::clustering}
\begin{figure}
	\includegraphics[width=0.94\linewidth]{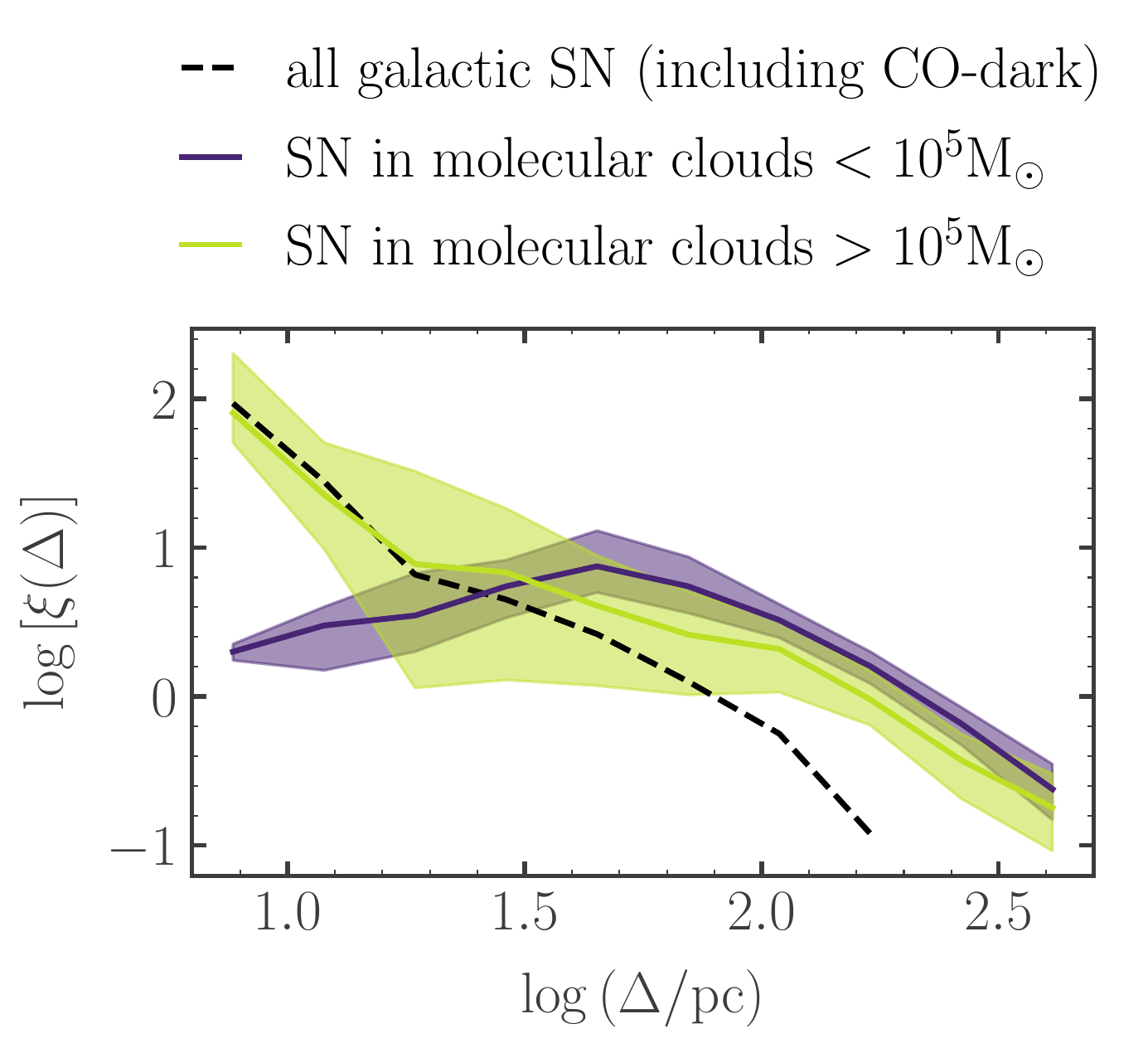}
	\caption{The two-point correlation function $\xi(\Delta)$ for supernova explosions as a function of their separation $\Delta$ over time intervals of 1~Myr, averaged over all times (solid lines). The shaded regions give the interquartile ranges over these times. The black dashed line is calculated for all supernovae across the entire simulation, while the green and purple lines represent the supernovae associated with star particles born in long-lived, high-mass and short-lived, low-mass molecular clouds, respectively. Massive and long-lived clouds account for supernova clustering on scales $\la 30~{\rm pc}$.}
	\label{Fig::SN-clustering}
\end{figure}

The concentration of star formation in the most massive and long-lived molecular clouds (Section~\ref{Sec::Eulerian-results}), fed by the fast accretion of new molecular gas (Section~\ref{Sec::competition-model}), drives the spatial and temporal clustering of supernovae. We demonstrate this in Figure~\ref{Fig::SN-clustering}, which shows the two-point correlation function $\xi(\Delta)$ of supernova explosions occurring in more and less massive clouds.
The black dashed line corresponds to the time-averaged median value of the two-point correlation function for all supernova explosions occurring across the simulated galaxy, within all time slices of 1~Myr over the 300~Myr simulation interval. This line indicates that on scales $\Delta \la 80~{\rm pc}$, the supernovae are \textit{more clustered} ($\xi>1$) than would be expected for a Poisson 
(uniform) distribution of objects across the galactic mid-plane, and on scales $\Delta > 80~{\rm pc}$ they are \textit{less clustered} ($\xi<1$).
The degree of clustering of supernovae rises steeply on small scales of $\la 30~{\rm pc}$, and Figure~\ref{Fig::SN-clustering} demonstrates that this clustering is accounted for almost exclusively by supernovae in clouds with masses $>10^5{\rm M}_\odot$ (green line). By contrast, the supernovae in lower-mass clouds (purple line) approach a random distribution at small scales. The clustering of supernovae on such small scales, approaching the softening length of $6$~pc used in our simulation, has been shown to enhance the momentum injected per supernova to the interstellar medium by a factor of at least $4$~\citep[e.g.][]{Gentry17}. This, in turn, increases the mass-loading of outflows~\citep{2018MNRAS.481.3325F} as well as the burstiness of star formation across galaxies~\citep{2021MNRAS.506.3882S}.

\begin{figure*}
  \includegraphics[width=\linewidth]{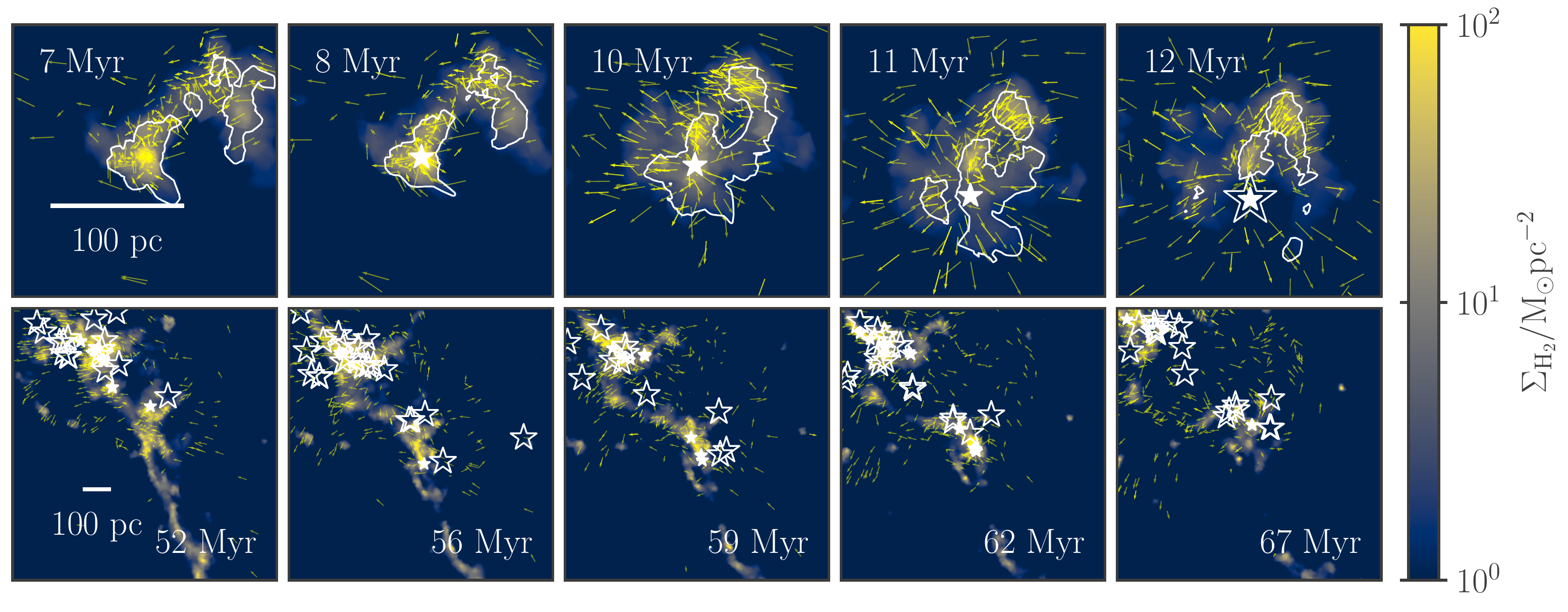}
  \caption{Examples of stellar feedback occurring in a low-mass, short-lived star-forming region ($\sim 5 \times 10^4~{\rm M}_\odot$, upper panels) and in a high-mass, long-lived star-forming region ($\sim 10^6~{\rm M}_\odot$, lower panels). Time runs from left to right across the page, with the region ages given in white. Star particles of age $<5~{\rm Myr}$ (emitting pre-supernova feedback) are indicated by filled white stars, while supernova explosions are indicated by open white stars. The velocities of the tracer particles that transit through the clouds are indicated by yellow arrows. The colorbar indicates the total molecular gas surface density, and white contours in the upper panel enclose regions of projected molecular hydrogen abundance $\Sigma_{\rm H_2}/\Sigma_{\rm g} \geq 0.3$. High-mass star-forming regions cannot be entirely destroyed by distributed pre-supernova feedback, and so are responsible for the clustering of supernovae on small scales in the simulation.}
  \label{Fig::tracer-fig}
\end{figure*}

Figure~\ref{Fig::tracer-fig} helps to visualise this intuitive result, comparing the time-evolving ${\rm H_2}$ column density of a low-mass cloud in our simulation ($\sim 5 \times 10^4~{\rm M}_\odot$, upper panels) to that of a high-mass cloud ($\sim 10^6~{\rm M}_\odot$, lower panels). White filled stars represent young star particles of age $<5~{\rm Myr}$, which are emitting significant quantities of ionising radiation, and empty stars denote supernova explosions. The yellow arrows show the velocities of gas tracer particles. In the low-mass cloud, star formation within a radius of $\sim 100~{\rm pc}$ is halted as the feedback from the young star particle destroys the entire molecular gas reservoir. By contrast, the high-mass cloud continues to accrete more molecular mass from the lower right-hand corner (inward-pointing yellow arrows), even as an entire cluster of supernovae are occurring in the top left-hand corner. Pre-supernova feedback destroys only small sections of the cloud, while new stars continue to form in other nearby sections. When massive stars eventually explode as supernovae, they are therefore surrounded by other nearby supernova explosions: a cluster of supernovae.

\section{Discussion} \label{Sec::discussion}
Here we discuss both the implications of our results in the context of the existing literature and the caveats associated with our calculations.

\subsection{Previous results for molecular cloud lifetimes}
Our demonstration that accretion-ejection balance is the key process in molecular cloud evolution helps resolve a number of problems in the literature regarding molecular cloud lifetimes. First, it helps explain why molecular cloud lifetimes computed in simulations of isolated GMCs tend to be too short compared to observations. In such isolated cloud simulations, molecular cloud lifetimes for observationally-motivated initial densities are typically $5-10$ Myr \citep[e.g.,][]{2018MNRAS.475.3511G,Grudic21a,2019MNRAS.489.1880H,2020MNRAS.497.3830F,Kim21a}, a factor of $2-5$ shorter than the $\approx 20-30$ Myr mean generally inferred from observations of nearby galaxies \citep[e.g.][]{Kawamura09, 2012ApJ...761...37M, 2019Natur.569..519K,Chevance20}. Our result provides a natural explanation for this discrepancy: simulations of isolated clouds are necessarily missing the accretion part of the accretion-ejection balance. With no fresh mass supplies, these simulations produce lifetimes closer to the Lagrangian lifetime of individual molecules than to the lifetimes of real molecular clouds.

Our results also help to explain the seeming contradiction between the presence of molecular clouds in the inter-arm regions of grand design spiral galaxies and a typical observed cloud lifetimes of $20-30$ Myr, much less than the inter-arm transit time. Our simulated dwarf spiral galaxy has an average gas density ($1-5$ cm$^{-3}$) and velocity dispersion ($\sim 10$ km s$^{-1}$), similar to the values found in the inter-arm regions of grand design spiral galaxies such as M51. Despite these conditions, and despite the short median survival time of an ${\rm H_2}$ molecule in our simulation ($\sim 4$~Myr, see Figures~\ref{Fig::moleculartime-vs-meanIH2-LCOGMCs} and~\ref{Fig::FB-IH2-vs-time}), we nonetheless find that massive molecular clouds in our simulation can survive for up to $90$~Myr. These long lifetimes are sustained by a high accretion rate of new molecular gas throughout the cloud lifetime: up to $4 \times 10^4~{\rm M}_\odot\,{\rm Myr}^{-1}$ for the longest-lived clouds. This accretion rate is an order of magnitude faster than predicted by~\cite{Koda09, Koda16a} for the comparable inter-arm environment in M51, and therefore provides an explanation for the existence of clouds in this inter-arm region that is not in contradiction with a much shorter typical molecule survival time ($\sim 4$~Myr) and a much shorter typical cloud survival time ($\approx 20-30$ Myr, see Figure~\ref{Fig::intro-CDFs}). Possible processes driving this fast accretion are converging flows due to supernova-driven bubbles or converging flows due to large-scale galactic-dynamical processes such as galactic shear. Pinpointing and describing this rate of accretion in our simulation will be the topic of a future paper.

\subsection{Caveats}
We conclude this discussion by pointing out two limitations of calculations, and considering their possible impact.

\subsubsection{Unresolved shielding of molecular gas}
The resolution of our simulation ($859~{\rm M_\odot}$, probing down to $\sim 2$~pc inside molecular clouds) presents a potential caveat to the result that the chemical survival time of ${\rm H_2}$ molecules is $4$~Myr in a dwarf spiral galaxy. Specifically, \cite{Joshi19} show that the ${\rm H_2}$ abundance derived using our chemical network and shielding prescription is converged only at a resolution of $0.2$~pc, due to the existence of sub-pc dense clumps of molecular gas.

In such clumps both the ratio of the H$_2$ formation and destruction rates and the amount of self-shielding would be enhanced compared to those we measure at our $\approx 2$ pc resolution, and as a result the value of $4$~Myr that we have derived for this lifetime may be an underestimate. This would also mean that the rate $\dot{M}_{\rm ej}$ of gas ejection from molecular clouds would be lowered relative to the value we have calculated. The rate $\dot{M}_{\rm accr}$ of gas accretion would be unaffected. However, there is circumstantial evidence that this cannot be a major effect. We have shown in Figure~\ref{Fig::observables} that the average ${\rm HI}$ and ${\rm H_2}$ gas densities and depletion times in our simulated galaxy are in good agreement with observations, as are the masses and sizes of its giant molecular clouds (Figure~\ref{Fig::cloud-diagnostics}). If the H$_2$ survival time were significantly longer, agreement with observations would worsen. This suggests that the rate limiting step in H$_2$ destruction is not so the timescale for chemical reactions, which we do not resolve, but the timescale for molecular material to be dynamically ejected from well-shielded clouds, which we do. Thus our failure to capture the details of the chemistry does minimal harm.
Given this discussion, however, we do caution that the ${\rm H_2}$ survival time of $4~{\rm Myr}$ depends on a number of factors, including the resolution of our simulation, the properties of the large-scale galactic environment, and the metallicity of the simulated gas reservoir (see next section). We have shown that this survival time is short, but great significance should not be attached to its exact value.

\subsubsection{Galactic-scale metallicity variations}
We have assumed a globally-invariant, Solar value for the gas-to-dust ratio in our simulation. However, the observed metallicity of NGC300 in fact varies between $0.8$ and $0.2$ times the Solar value, decreasing monotonically with galactocentric radius~\citep{2009ApJ...700..309B}. According to our competition model for the evolution of molecular clouds, lowering the gas-phase metallicity in the outer regions of the galactic disc may have several different (competing) effects. Reducing the metallicity will reduce the shielding of ${\rm H_2}$ from the ISRF, and so reduce the distance from young stars at which ${\rm H_2}$ is ionised as it is ejected by early stellar feedback. This will cause an increase in the rate $\dot{M}_{\rm ej}$ of ${\rm H_2}$ ejection from molecular clouds. However, lowered metallicity will also reduce the degree of cooling and star formation in the dense gas, causing a decrease in $\dot{M}_{\rm ej}$. The effect of lowered metallicity on the mass accretion rate $\dot{M}_{\rm accr}$ is more straight-forward: (1) the dust-catalysed formation of ${\rm H_2}$ from atomic hydrogen will be slower, and (2) gas parcels will need to reach higher densities before becoming shielded from the ISRF. Both of these effects will act to decrease the rate of ${\rm H_2}$ accretion onto molecular clouds.

As such, it is difficult to predict exactly how the inclusion of live metal-enrichment will affect our results, except to note that the environmentally-dependent metallicity will likely be a variable that plays a significant role in setting $\dot{M}_{\rm ej}$. Further studies that vary the metallicity and control for galaxy morphology are necessary to determine the functional form of this dependence.

\section{Conclusions} \label{Sec::conclusions}
In Section~\ref{Sec::Introduction} we posed two questions: (1) what is the chemical lifetime of ${\rm H_2}$ in the interstellar medium of galaxies and how does it relate to the molecular cloud lifetime? And (2) how can we reconcile the long observed lifetimes of giant molecular clouds~\citep[e.g.][]{Koda09} and the short Lagrangian residence times of gas in the star-forming state~\citep[e.g.][]{Semenov17}? In this work, we have used passive gas tracer particles in an {\sc Arepo} simulation of a dwarf spiral galaxy to arrive at the following answers:
\begin{enumerate}
	\item The typical survival time of ${\rm H_2}$ molecules is $4~{\rm Myr}$ in our simulation, independent of the lifetime of the host molecular cloud, which has a much longer median value of $\sim 25~{\rm Myr}$. The rapid destruction of ${\rm H_2}$ molecules is driven by early stellar feedback from HII regions.
	\item  The reason that molecular clouds are able to survive much longer than their constituent molecules is that the molecular cloud lifetime is determined by the competition between the feedback-driven ejection of molecular gas, and the accretion of new molecular gas from the large-scale galactic environment. This competition means that long-lived and massive molecular clouds can be sustained given a sufficiently-fast accretion rate. In our simulation, a value of $\dot{M}_{\rm accr} \sim 4 \times 10^4\,{\rm M}_\odot\,{\rm Myr}^{-1}$ can feed a molecular cloud for $90$~Myr, and up to a peak mass of $10^6~{\rm M}_\odot$.
\end{enumerate}

Our answers have important implications for the dynamics of gas on both cloud and galactic scales. In particular, we find that
\begin{enumerate}
    \item The time spent by gas in the ${\rm H_2}$-dominated state is much shorter than the time-spent in the ${\rm H_2}$-poor state, in qualitative agreement with the results of~\cite{Semenov17}. A given Lagrangian parcel of gas spends most of its life transiting between molecular clouds, with only brief passages through clouds.
    \item A partial exception to this is that gas can sometimes become `trapped' around higher-mass molecular clouds, cycling back to a dense, ${\rm H_2}$-dominated state faster and more frequently than occurs for gas parcels passing through low-mass clouds. This `trapping' raises the integrated star formation efficiency in massive, long-lived molecular clouds up to four times higher than that in low-mass, short-lived clouds.
    \item Despite their small number, massive and long-lived molecular clouds account for a not insignificant amount of the total star formation budget of a galaxy. Moreover, they drive the clustering of supernova feedback on sub-cloud scales, which is in turn a key driver of galactic outflows.
\end{enumerate}

In an upcoming paper, we will investigate the galactic-scale physics driving the fast accretion of new gas onto massive, long-lived molecular clouds, and so driving their higher star formation efficiencies and levels of supernova clustering.

\section*{Acknowledgements}
We are very grateful to Shy Genel, Greg Bryan, Blakesley Burkhart, Enrique V\'{a}zquez-Semadeni, Charles Lada, Shyam Menon and Munan Gong for helpful discussions.
We thank Volker Springel for providing us access to Arepo. SMRJ is supported by Harvard University through the Institute for Theory and Computation Fellowship. Support for VS was provided by NASA through the NASA Hubble Fellowship grant HST-HF2-51445.001-A awarded by the Space Telescope Science Institute, which is operated by the Association of Universities for Research in Astronomy, Inc., for NASA, under contract NAS5-26555, and by Harvard University through the Institute for Theory and Computation Fellowship. MRK acknowledges support from the Australian Research Council through Laurate Fellowship FL220100020. The work was undertaken with the assistance of resources and services from the National Computational Infrastructure (NCI; award jh2), which is supported by the Australian Government. 

\section*{Data Availability Statement}
The data underlying this article are available in the article and in its online supplementary material.



\bibliographystyle{mnras}
\bibliography{bibliography} 



\appendix
\section{Convergence of the initial tracer mass distribution} \label{App:tracer-error}
\begin{figure}
	\includegraphics[width=\linewidth]{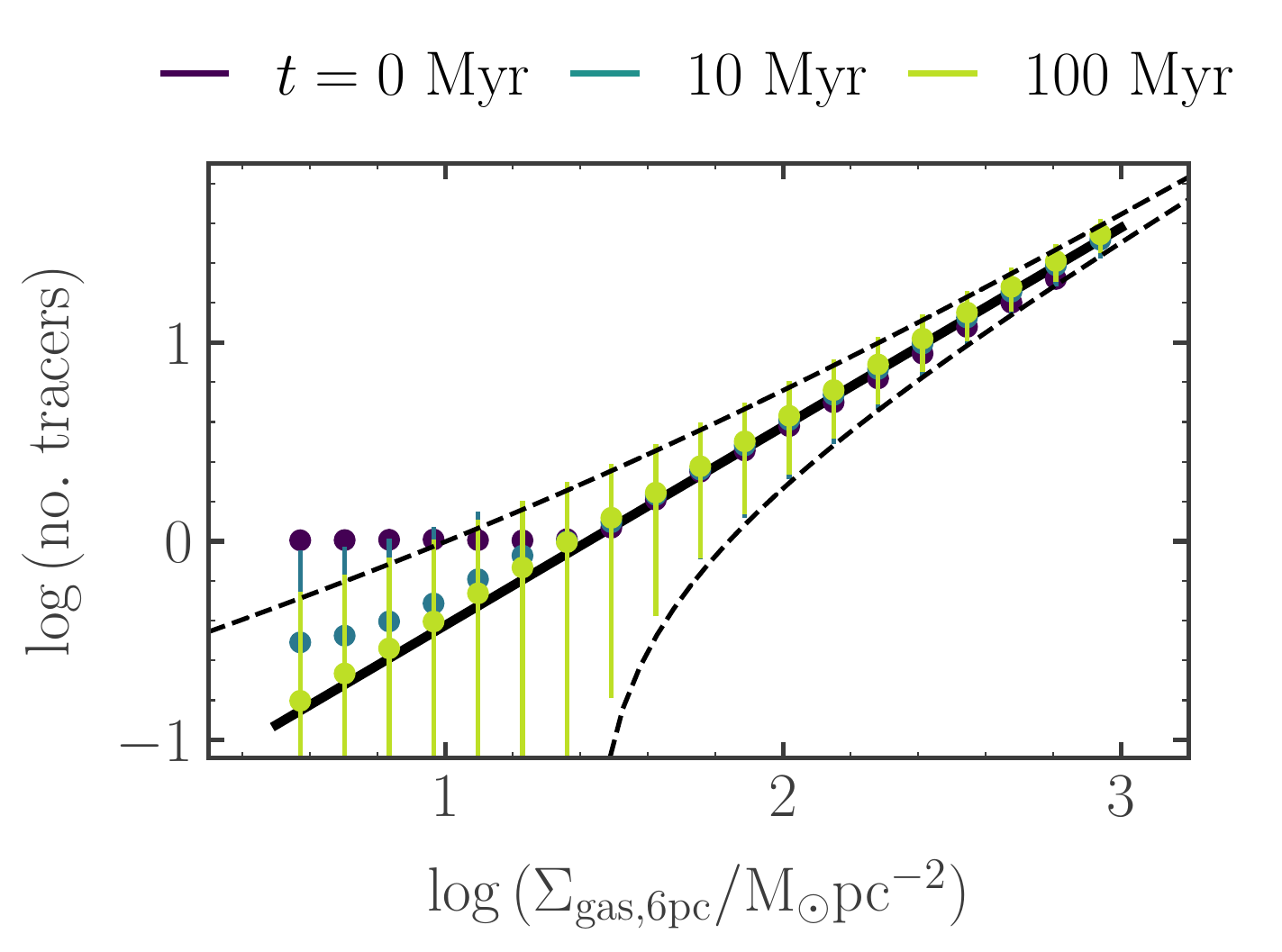}
	\caption{Mean and standard deviation of the tracer number as a function of gas surface density in 2D columns of area $(6~{\rm pc})^2$ through the entire mid-plane of the simulated galaxy disc, at simulation times of $0$~Myr, $10$~Myr and $100$~Myr. The pixel size corresponds to the native resolution of the simulation, at which molecular clouds are identified and analysed in two dimensions. The thick black line corresponds to one tracer particle per $450~{\rm M}_\odot$, while the dashed black lines show a range of $\pm\sqrt{\rm no.~tracers}$ around this, the error expected from Poisson sampling for $\mathrm{no.~tracers}\gg 1$.}
	\label{Fig::poisson-error}
\end{figure}

In Figure~\ref{Fig::poisson-error}, we demonstrate that the initial, non-uniform effective mass distribution of tracer particles in our simulation converges to a uniform distribution on a time-scale of $<100$~Myr. As noted in Section~\ref{Sec::sims::tracers}, we assign one tracer particle per gas cell in our initial condition, and the initial spread of gas cell masses is over an order of magnitude. This means that the initial spread of effective tracer particle masses is over an order of magnitude.

In order to demonstrate that this initial non-uniform distribution rapidly approaches one that samples uniformly in mass, in Figure~\ref{Fig::poisson-error} we plot the number of tracer particles as a function of gas surface density computed in 2D columns through the galactic mid-plane, at the native resolution of the simulation ($6$~pc). The black solid line has slope unity, and shows the expected relationship for a uniform distribution of effective tracer particle masses, while the dashed lines denote the Poisson error expected due to the Monte Carlo exchange of a small number of tracer particles between gas cells. The plot shows that, after a period of $100$~Myr (green points and error bars), both the number of tracer particles per pixel, and the error in this number, conform to the expectation for a uniform distribution of effective tracer particle masses.

\section{Chemical post-processing} \label{App::chem-postproc}
As noted in Section~\ref{Sec::sim-props}, our CO-luminous star-forming regions are identified using two-dimensional maps of the CO-bright molecular gas column density, $\Sigma_{\rm H_2, CO}$. To calculate this column density, we post-process the simulation output using the {\sc Despotic} model for astrochemistry and radiative transfer~\citep{Krumholz13a, Krumholz14}, following the method outlined in \citet{Fujimoto19} except as noted below; we summarise the procedure here, and refer readers to \citeauthor{Fujimoto19} for further details. The self- and dust-shielding of CO molecules from the ambient UV radiation field, and the effects of non-LTE excitation and varying CO line optical depth, cannot be accurately computed during run-time at the mass resolution of our simulation. Within {\sc Despotic}, the escape probability formalism is applied to compute the CO line emission from each gas cell according to its hydrogen atom number density $n_{\rm H}$, column density $N_{\rm H}$ and virial parameter $\alpha_{\rm vir}$, assuming that the cells are approximately spherical. In practice, the line luminosity varies smoothly with the variables $n_{\rm H}$, $N_{\rm H}$, and $\alpha_{\rm vir}$. We therefore interpolate over a grid of pre-calculated models at regularly-spaced logarithmic intervals in these variables to reduce computational cost. The hydrogen column density is estimated via the local approximation of~\cite{Safranek-Shrader+17} as $N_{\rm H}=\lambda_{\rm J} n_{\rm H}$, where $\lambda_{\rm J}=(\pi c_s^2/G\rho)^{1/2}$ is the Jeans length, with an upper limit of $T=40~{\rm K}$ on the gas cell temperature. The virial parameter is calculated from the turbulent velocity dispersion of each gas cell according to~\citet{MacLaren1988} and \citet{BertoldiMcKee1992}. The line emission is self-consistently coupled to the chemical and thermal evolution of the gas, including carbon and oxygen chemistry~\citep{Gong17}, gas heating by cosmic rays and the grain photo-electric effect, line cooling due to ${\rm C}^+$, ${\rm C}$, ${\rm O}$ and ${\rm CO}$ and thermal exchange between dust and gas. We match the ISRF strength and cosmic ray ionisation rate to the values used in our live chemistry.

Having calculated values of the CO line luminosity for each simulated gas cell, we compute the CO-bright molecular hydrogen surface density as
\begin{equation}
\begin{split}
\Sigma_{\rm H_2, CO}[{\rm M}_\odot{\rm pc}^{-2}] = &\frac{2.3 \times 10^{-29}{\rm M}_\odot({\rm erg~s}^{-1})^{-1}}{m_{\rm H}[{\rm M}_\odot]} \\
&\times \int^{\infty}_{-\infty}{\dd z^\prime \rho_{\rm g}(z^\prime) L_{\rm CO}[{\rm erg~s}^{-1}~{\rm H~atom}^{-1}]},
\end{split}
\end{equation}
where $\rho_{\rm g}(z)$ is the total gas volume density in ${\rm M}_\odot~{\rm pc}^{-3}$ at a distance $z$ (in ${\rm pc}$) from the galactic mid-plane. The factor of $2.3 \times 10^{-29}~{\rm M}_\odot~({\rm erg~s}^{-1})^{-1}$ combines the mass-to-luminosity conversion factor $\alpha_{\rm CO}=4.3~{\rm M}_\odot{\rm pc}^{-2}({\rm K}~{\rm kms}^{-1})^{-1}$ of~\cite{Bolatto13} with the line-luminosity conversion factor $5.31 \times 10^{-30}({\rm K~km~s}^{-1}{\rm pc}^2)/({\rm erg~s}^{-1})$ for the CO $J=1\rightarrow 0$ transition at redshift $z=0$~\citep{SolomonVandenBout05}. The physical meaning of $\Sigma_\mathrm{H_2,CO}$ is simply that it is the H$_2$ surface density that one would infer from an observation in the CO $J=1\to 0$ line assuming a fixed conversion factor $\alpha_{\rm CO}=4.3~{\rm M}_\odot{\rm pc}^{-2}({\rm K}~{\rm kms}^{-1})^{-1}$.


\bsp	
\label{lastpage}
\end{document}